\documentclass[hyper]{JHEP3}
\usepackage{epsfig}
\usepackage{cite}
\def\beq{\begin{equation}}
\def\eeq{\end{equation}}
\def\beeq{\begin{eqnarray}}
\def\eeeq{\end{eqnarray}}

\def\nn{\nonumber}
\def\as{\alpha_{\mathrm{S}}}

\def\res{{\rm res.}}
\def\ms{${\overline {\rm MS}}$}

\def\vec#1{\mbox{\boldmath $#1$}}
\def\vecex#1{\mbox{\boldmath\scriptsize $#1$}}
\def\tC{\widetilde{C}}
\def\etamax{\eta_{\mbox{\scriptsize max}}}

\def\rs{\sqrt{s}}

\title{Resummation of transverse energy in vector boson and Higgs boson production at hadron colliders}

\author{Andreas Papaefstathiou$^a$, Jennifer M.\ Smillie$^b$ and Bryan R.\ Webber$^a$\\
        $^a$Cavendish Laboratory, J.J.\ Thomson Avenue, Cambridge, UK\\
        $^b$Department of Physics and Astronomy, University College London, WC1E 6BT, UK\\
        E-mail: \email {andreas@hep.phy.cam.ac.uk}, \email{smillie@hep.ucl.ac.uk}, \email{webber@hep.phy.cam.ac.uk}
        }

\received{\today}               
\accepted{\today}               

\preprint{Cavendish-HEP-10/01\\MCnet/10/01}

\abstract{
We compute the resummed hadronic transverse energy ($E_T$)
distribution due to initial-state QCD radiation in vector boson and
Higgs boson production at hadron colliders.  The resummed exponent,
parton distributions and coefficient functions are treated consistently to
next-to-leading order.
The results are matched to fixed-order calculations at large $E_T$ and
compared with parton-shower Monte Carlo predictions at Tevatron and
LHC energies.
}

\keywords{Hadronic Colliders, QCD Phenomenology}

\begin{document} 

\section{Introduction}
\label{sec:intro}
The QCD radiation from incoming partons forms an inescapable component
of the final state in all hard scattering processes at hadron
colliders.  This radiation leads to hadron formation that complicates
the interpretation of events in a number of ways: by
generating extra jets, by contaminating other jets, by modifying event
shapes and global observables, and by changing
the distributions of the products of the hard process.  This
last effect has been studied in great detail for the processes of
electroweak boson production, with the result that the transverse
momentum and rapidity distributions of W, Z and Higgs bosons at the
Tevatron and LHC are predicted with good
precision.\footnote{See~\cite{Bozzi:2007pn,Bozzi:2008bb,Mantry:2009qz}
  and references therein.}  The predictions
for the transverse momentum ($q_T$) distributions in particular include
resummation of terms enhanced at small $q_T$ to all orders in $\as$,
matched with fixed-order calculations at higher $q_T$ values.
The transverse momentum of the boson arises (neglecting
the small intrinsic transverse momenta of the partons in the
colliding hadrons) from its recoil against the transverse momenta of
the radiated partons: $q_T=|\vec q_T|$ where
\beq\label{eq:qt}
\vec q_T = -\sum_i\vec p_{Ti}\;.
\eeq
The resummation of enhanced terms therefore requires a sum over
emissions $i$ subject to the constraint (\ref{eq:qt}), which is most
conveniently carried out in the transverse space of the impact
parameter $\vec b$ Fourier conjugate to $\vec q_T$:
\beq\label{eq:deltaqt}
\delta(\vec q_T +\sum\vec p_{Ti}) = \frac 1{(2\pi)^2}\int d^2\vec b\,{\rm e}^{i\vecex
  q_T\cdot\vecex b}\prod_i {\rm e}^{i\vecex p_{Ti}\cdot\vecex b}\;.
\eeq
One then finds that the cumulative distribution in $b=|\vec b|$
contains terms of the form $\as^n\ln^p(Qb)$, where $Q$ is the
scale of the hard process, set in this case by the mass of the
electroweak boson, and $p\leq 2n$.  These terms, which spoil the convergence
of the perturbation series at large $b$, corresponding to small $q_T$,
are found to exponentiate~\cite{Dokshitzer:1980hw,Parisi:1979se,Curci:1979bg,Bassetto:1980nt,Kodaira:1982nh,Collins:1984kg}: that is, they can be assembled into an
exponential function of terms that are limited to $p\leq n+1$.
This resummation procedure improves the convergence of the perturbation
series at large values of $b$ and hence allows one to extend predictions of the
$q_T$ distribution to smaller values. 

Together with its vector transverse momentum $\vec p_{Ti}$, every
emission generates a contribution to the total hadronic transverse energy
of the final state, $E_T$, which, neglecting parton masses, is given by
\beq\label{eq:Et}
E_T = \sum_i|\vec p_{Ti}|\;.
\eeq
To first order in $\as$ (0 or 1 emissions) this quantity coincides
with $q_T$, but they differ in higher orders.  In particular, at small
$q_T$ there is the possibility of vectorial cancellation between the
contributions of different emissions, whereas this cannot happen for
the scalar $E_T$.  Thus the distribution of $E_T$ vanishes faster at
the origin, and its peak is pushed to higher values.  To resum these
contributions at small $E_T$, one should perform a one-dimensional
Fourier transformation and work in terms of a `transverse time'
variable $\tau$ conjugate to $E_T$:
\beq\label{eq:deltaEt}
\delta(E_T -\sum|\vec p_{Ti}|) = \frac 1{2\pi}\int d\tau\,{\rm e}^{-iE_T\tau}\prod_i {\rm e}^{i|\vecex p_{Ti}|\tau}\;.
\eeq
Since the matrix elements involved are the same, one finds a similar
pattern of enhanced terms at large $\tau$ as was the case for large
$b$: terms of the form $\as^n\ln^p(Q\tau)$ with $p\leq 2n$, which
arise from an exponential function of terms with $p\leq n+1$.
Evaluation of the exponent to a certain level of precision
(leading-logarithmic, LL, for $p=n+1$, next-to-leading, NLL, for
$p=n$, etc.) resums a corresponding class of enhanced terms and
extends the validity of predictions to lower values of $E_T$.

The resummation of $E_T$ in this way has received little attention
since the first papers on this topic, over 20 years
ago~\cite{Halzen:1982cb,Davies:1983di,Altarelli:1986zk}.  This is surprising, as
most of the effects of QCD radiation from incoming partons mentioned
above depend on this variable rather than $q_T$.  A possible reason is
that, unlike $q_T$, $E_T$ also receives an important contribution from
the so-called underlying event, which is thought to arise from
secondary interactions between spectator partons.  At present this can
only be estimated from Monte Carlo simulations that include multiple
parton interactions (MPI).  Nevertheless it is worthwhile to predict
as accurately as possible the component coming from the primary
interaction, which carries important information about the hard
process.  For example, we expect the $E_T$ distributions in Higgs and
vector boson production to be different, as they involve
primarily gluon-gluon and quark-antiquark annihilation, respectively.
Accurate estimates of the primary $E_T$ distribution are also
important for improving the modelling of the underlying event.

In the present paper we extend the resummation of $E_T$ in vector
boson production to next-to-leading order (NLO) in the resummed exponent,
parton distributions and coefficient functions, and present for the
first time the corresponding predictions for Higgs boson production.  
In Section~\ref{sec:method} the resummation procedure is reviewed and extended to
NLO; results on the resummed component are presented in Sect.~\ref{sec:resum}.
This component alone is not expected to describe the region of larger $E_T$
values, of the order of the boson mass; in Sect.~\ref{sec:match} we
describe and apply a simple procedure for including the
unresummed component at order $\as$.  Section~\ref{sec:MC} presents
$E_T$ distributions generated using  the parton shower Monte Carlo
programs {\tt HERWIG}~\cite{Corcella:2000bw} and
{\tt Herwig++}~\cite{Bahr:2008pv}, which are compared with the
analytical results and used to estimate of the effects of hadronization and
the underlying event.  Our conclusions are summarized in
Sect.~\ref{sec:conc}.  Appendix A gives mathematical details of a
comparison between the resummation of the transverse energy $E_T$
and transverse momentum $q_T$ and Appendix B shows results for the
LHC at lower centre-of-mass energy.

\section{Resummation method}\label{sec:method}
\subsection{General procedure}
Here we generalize the results of ref.~\cite{Davies:1983di} to NLO resummation.
The resummed component of the transverse energy distribution in the
process $h_1h_2\to FX$ at scale $Q$ takes the form
\beeq
\label{resgen}
\left[ \frac{d\sigma_{F}}{dQ^2\;dE_T} \right]_{\res} &=& \frac 1{2\pi}\sum_{a,b}
\int_0^1 dx_1 \int_0^1 dx_2 \int_{-\infty}^{+\infty} d\tau \; {\rm e}^{-i\tau E_T} 
\;f_{a/h_1}(x_1,\mu) \; f_{b/h_2}(x_2,\mu) \nn \\
&\cdot& W_{ab}^{F}(x_1 x_2 s; Q, \tau,\mu)
\eeeq
where $f_{a/h}(x,\mu)$ is the parton distribution function (PDF) of
parton $a$ in hadron $h$ at factorization scale $\mu$, taken to be
the same as the renormalization scale here.  In what follows we use
the \ms\ renormalization scheme.   As mentioned earlier,  to take into
account the constraint that the transverse energies of emitted partons
should sum to $E_T$,  the resummation procedure is carried out in the
domain that is Fourier conjugate to $E_T$, using
Eq.~(\ref{eq:deltaEt}).  The transverse energy distribution
(\ref{resgen}) is thus obtained by performing the inverse Fourier
transformation with respect to the transverse time, $\tau$.  The
factor $W_{ab}^{F}$ is the perturbative and  process-dependent
partonic cross section that embodies the all-order  resummation of the
large logarithms $\ln (Q\tau)$.  Since $\tau$ is  conjugate to $E_T$,
the limit $E_T\ll Q$ corresponds to $Q\tau \gg 1$.

As in the case of transverse momentum resummation~\cite{Catani:2000vq},
the resummed partonic cross section can be written in the
following universal form:
\beeq
\label{eq:Wab}
W_{ab}^{F}(s; Q, \tau,\mu) &=& \sum_c \int_0^1 dz_1 \int_0^1 dz_2 
\; C_{ca}(\as(\mu), z_1;\tau,\mu) \; C_{{\bar c}b}(\as(\mu), z_2;\tau,\mu)
\; \delta(Q^2 - z_1 z_2 s) \nn\\
&\cdot& \sigma_{c{\bar c}}^F(Q,\as(Q)) \;S_c(Q,\tau) \;\;.
\eeeq
Here $\sigma_{c{\bar c}}^F$ is the cross section
for the partonic subprocess
$c + {\bar c} \to F$, where $c,{\bar c}=q,{\bar q}$ (the quark $q_f$ and
the antiquark ${\bar q}_{f'}$ can possibly have different flavours $f,f'$)
or $c,{\bar c}=g,g$. The term $S_c(Q,\tau)$ is the quark $(c=q)$ or
gluon $(c=g)$ Sudakov form factor.  In the case of $E_T$ resummation,
this takes the form~\cite{Davies:1983di,Altarelli:1986zk}
\beq
\label{formfact}
S_c(Q,\tau) = \exp \left\{-2\int_0^Q \frac{dq}q 
\left[ 2A_c(\as(q)) \;\ln \frac{Q}{q} + B_c(\as(q)) \right] 
\left(1-{\rm e}^{iq\tau}\right)\right\} \;, 
\eeq
with $c=q$ or $g$. The functions $A_c(\as), B_c(\as)$, as well as the 
coefficient functions $C_{ab}$ in Eq.~(\ref{eq:Wab}), contain no
$\ln (Q\tau)$ terms and are perturbatively computable as power
expansions with constant coefficients:
\beeq
\label{aexp}
A_c(\as) &=& \sum_{n=1}^\infty \left( \frac{\as}{\pi} \right)^n A_c^{(n)} 
\;\;, \\
\label{bexp}
B_c(\as) &= &\sum_{n=1}^\infty \left( \frac{\as}{\pi} \right)^n B_c^{(n)}
\;\;, \\
\label{cexp}
C_{ab}(\as,z) &=& \delta_{ab} \,\delta(1-z) + 
\sum_{n=1}^\infty \left( \frac{\as}{\pi} \right)^n C_{ab}^{(n)}(z) \;\;.
\eeeq
Thus a calculation to NLO in $\as$ involves the coefficients
$A_c^{(1)}$,  $A_c^{(2)}$,  $B_c^{(1)}$,  $B_c^{(2)}$ and  $C_{ab}^{(1)}$.
All these quantities are known for both the quark and gluon form
factors and associated coefficient functions.  Knowledge of the
coefficients $A^{(1)}$ leads to the resummation of
the leading logarithmic (LL) contributions at small $E_T$,
which in the differential distribution are of the form
$\as^n\ln^p(Q/E_T)/E_T$ where $p=2n-1$.
The coefficients $B^{(1)}$ give the next-to-leading logarithmic (NLL)
terms with $p=2n-2$, $A^{(2)}$ and $C^{(1)}$ give the 
next-to-next-to-leading logarithmic (N$^2$LL) terms with $p=2n-3$,
and $B^{(2)}$ gives the N$^3$LL terms with $p=2n-4$. 
With knowledge of all these terms, the first term neglected
in the resummed part of the distribution is of order $\as^3\ln(Q/E_T)/E_T$.

In general the coefficient functions in Eq.~(\ref{eq:Wab}) contain logarithms of
$\mu\tau$, which are eliminated by a suitable choice of factorization scale.
To find the optimal factorization scale, we note that, to NLL accuracy,
\beq\label{eq:nllint}
\int_0^Q\frac{dq}q \ln^p q \left(1-{\rm e}^{iq\tau}\right)
\simeq \int_{i\tau_0/\tau}^Q\frac{dq}q \ln^p q \;,
\eeq
where $\tau_0 = \exp(-\gamma_E)=0.56146\ldots$, $\gamma_E$ being the
Euler-Mascheroni constant. Therefore the effective lower limit of the
soft resummation is $i\tau_0/\tau$, and the parton distributions and
coefficient functions should be evaluated at this scale.  However,
evaluation of parton distribution functions at an imaginary scale
using the standard parametrizations is not feasible.  We avoid this by
noting that
\beq
 f_{a/h}(x,q') = \sum_b\int_x^1 \frac{dz}z U_{ab}(z;q',q) f_{b/h}(x/z,q)
\eeq
where $U_{ab}$ is the DGLAP evolution operator.  Therefore
\beq\label{eq:fabi}
 f_{a/h}(x,i\mu) = \int_x^1 \frac{dz}z U_{ab}(z;i\mu,\mu) f_{b/h}(x/z,\mu)
\eeq
where the evolution operator $U_{ab}(z;i\mu,\mu)$ is given to NLO by
\beq\label{eq:Uabi}
U_{ab}(z;i\mu,\mu) = \delta_{ab} + \frac i2 \as(\mu)\,P_{ab}(z)\;,
\eeq
$P_{ab}(z)$ being the leading-order DGLAP splitting function.
Similarly, in the coefficient functions we can write $\as(i\mu)$ in terms of  $\as(\mu)$
using the definition of the running coupling:
\beq
\int_{\mu}^{i\mu}\frac{d\as}{\beta(\as)} = 2\int_{\mu}^{i\mu}\frac{dq}q = i\pi
\eeq
where $\beta(\as)=-b\as^2 +{\cal O}(\as^3)$, so that
\beq
\as(i\mu) = \as(\mu) -i\pi b[\as(\mu)]^2 +{\cal O}(\as^3)\;.
\eeq
Furthermore, as the expressions (\ref{resgen}) and (\ref{eq:Wab}) are
convolutions, we can transfer the extra terms from (\ref{eq:fabi})
into the coefficient functions to obtain
\beeq
\label{eq:Wfin}
W_{ab}^{F}(s; Q, \tau) &=& \sum_c \int_0^1 dz_1 \int_0^1 dz_2 
\; \tC_{ca}(\as(\tau_0/\tau), z_1) \; \tC_{{\bar c}b}(\as(\tau_0/\tau), z_2)
\; \delta(Q^2 - z_1 z_2 s) \nn\\
&\cdot& \sigma_{c{\bar c}}^F(Q,\as(Q)) \;S_c(Q,\tau)
\eeeq
where
\beq\label{eq:tCca}
\tC_{ca}(\as(\mu),z) = \sum_d \int_z^1 \frac{dz'}{z'}C_{cd}(\as(i\mu),z/z')\,U_{da}(z';i\mu,\mu)\;.
\eeq
Now the lowest-order coefficient function is of the form
\beq\label{eq:tC0}
\tC_{ca}^{(0)}(z) = C_{ca}^{(0)}(z) = \delta_{ca}\delta(1-z)
\eeq
and therefore
\beq\label{eq:tC1}
\tC_{ca}^{(1)}(z) = C_{ca}^{(1)}(z) +i\frac{\pi}2 P_{ca}(z)\;.
\eeq

Putting everything together, we have
\beq\label{eq:resSR}
\left[ \frac{d\sigma_{F}}{dQ^2\;dE_T} \right]_{\res}  = \frac 1{2\pi s}\sum_{c}
\int_{-\infty}^{+\infty} d\tau \; {\rm e}^{-i\tau E_T} S_c(Q,\tau)\;R_c(s;Q,\tau)
\;\sigma_{c{\bar c}}^F(Q,\as(Q))
\eeq
where, taking all PDFs and coefficient functions to be evaluated at scale $\mu=\tau_0/\tau$, 
\beq\label{eq:Rcdef}
R_c(s;Q,\tau) =\sum_{a,b}
\int_0^1 \frac{dx_1}{x_1} \frac{dx_2}{x_2} \frac{dz_1}{z_1} 
f_{a/h_1}(x_1)\,f_{b/h_2}(x_2)\,\tC_{ca}(z_1)\,\tC_{\bar cb}\left(\frac{Q^2}{z_1x_1x_2 s}\right)\;.
\eeq
To write (\ref{eq:resSR}) as an integral over $\tau>0$ only, we note from (\ref{eq:fabi}) and (\ref{eq:Uabi})
that when $\tau\to -\tau$, to NLO the real parts of $f_{a/h_1}$ and $f_{b/h_2}$ are unchanged but
the imaginary parts change sign.  All other changes in (\ref{eq:Rcdef}) are beyond NLO.  Thus, writing
\beq
R_c= R_c^{(R)}  + iR_c^{(I)}\;,
\eeq
$R_c^{(R)}$ is symmetric with respect to $\tau$ and $R_c^{(I)}$ is antisymmetric.
Defining
\beeq\label{eq:Fcs}
F_c^{(R)}(Q,\tau) &=& 2\int_0^Q \frac{dq}q 
\left[ 2A_c(\as(q)) \;\ln \frac{Q}{q} + B_c(\as(q)) \right] 
\left(1-\cos q\tau\right)\;,\nn\\
F_c^{(I)}(Q,\tau) &=& 2\int_0^Q \frac{dq}q 
\left[ 2A_c(\as(q)) \;\ln \frac{Q}{q} + B_c(\as(q)) \right] 
\sin q\tau
\eeeq
we therefore obtain
\beeq\label{eq:resF}
\left[ \frac{d\sigma_F}{dQ^2\;dE_T} \right]_{\res} &=&
\frac 1{\pi s}\sum_{c} \int_{0}^{\infty}
 d\tau \; {\rm e}^{-F_c^{(R)}(Q,\tau)} \Bigl[
 R_c^{(R)}(s;Q,\tau)\cos\{F_c^{(I)}(Q,\tau)-\tau E_T\}\nn\\
&&-R_c^{(I)}(s;Q,\tau)\sin\{F_c^{(I)}(Q,\tau)-\tau E_T\}\Bigr]
\;\sigma_{c{\bar c}}^F(Q,\as(Q))
\eeeq
where, inserting (\ref{eq:tC0}) and (\ref{eq:tC1}) in (\ref{eq:Rcdef})
and defining $\xi=Q^2/s$, we have to NLO
\beeq\label{eq:RcNLO}
&&R_c^{(R)}(s;Q,\tau) = R_c^{(R)}(\xi=Q^2/s,\tau) \nn\\
&&=\int \frac{dx_1}{x_1} \frac{dx_2}{x_2}
\Bigl\{f_{c/h_1}(x_1)f_{\bar c/h_2}(x_2)+\frac{\as}{\pi}\sum_a\Bigl[
f_{a/h_1}(x_1)f_{\bar c/h_2}(x_2)C_{ca}^{(1)}\left(\frac{\xi}{x_1x_2}\right)\nn\\
&&\quad +f_{c/h_1}(x_1)f_{a/h_2}(x_2)C_{\bar ca}^{(1)}\left(\frac{\xi}{x_1x_2}\right)
\Bigr]\Bigr\}\;,\nn\\
&&R_c^{(I)}(s;Q,\tau) = R_c^{(I)}(\xi=Q^2/s,\tau) \nn\\
&&=\frac{\as}2\sum_a\int \frac{dx_1}{x_1} \frac{dx_2}{x_2}
\Bigl[f_{a/h_1}(x_1)f_{\bar c/h_2}(x_2)P_{ca}\left(\frac{\xi}{x_1x_2}\right)\nn\\
&&\quad +f_{c/h_1}(x_1)f_{a/h_2}(x_2)P_{\bar ca}\left(\frac{\xi}{x_1x_2}\right)
\Bigr]\;.
\eeeq
It will be more useful to write, for example,
\beeq\label{eq:ffPz}
&&\quad \int \frac{dx_1}{x_1} \frac{dx_2}{x_2}
f_{a/h_1}(x_1)f_{\bar c/h_2}(x_2)P_{ca}\left(\frac{\xi}{x_1x_2}\right)\nn\\
&&=\int \frac{dx_1}{x_1} \frac{dx_2}{x_2}dz\,\delta\left(z-\frac{\xi}{x_1x_2}\right)
f_{a/h_1}(x_1)f_{\bar c/h_2}(x_2)P_{ca}(z)\nn\\
&&=\int \frac{dx_1}{x_1} \frac{dz}{z} f_{a/h_1}(x_1)f_{\bar c/h_2}\left(\frac{\xi}{zx_1}\right)P_{ca}(z)\;.
\eeeq
This makes it more straightforward to interpret the +-prescription, which appears in some splitting functions, as
\beeq\label{eq:ffPplus}
&& \int \frac{dx_1}{x_1} \frac{dz}{z} f_{a/h_1}(x_1)f_{\bar c/h_2}\left(\frac{\xi}{zx_1}\right)P(z)_+\nn\\
&=&\int \frac{dx_1}{x_1} f_{a/h_1}(x_1)\int_0^1 dz\left[\frac 1z f_{\bar c/h_2}\left(\frac{\xi}
{zx_1}\right) - f_{\bar c/h_2}\left(\frac{\xi}{x_1}\right) \right] P(z)\nn\\
&=&\int_{\xi}^1 \frac{dx_1}{x_1} f_{a/h_1}(x_1)\int_{\xi/x_1}^1 dz\,\left[\frac 1z
f_{\bar c/h_2}\left(\frac{\xi}{zx_1}\right) - f_{\bar c/h_2}\left(\frac{\xi}{x_1}\right) \right] P(z)\nn\\
&&\quad -\int_{\xi}^1 \frac{dx_1}{x_1} f_{a/h_1}(x_1)f_{\bar
  c/h_2}\left(\frac{\xi}{x_1}\right) \int_0^{\xi/x_1} dz\,P(z)\;.
\eeeq

We show in Appendix A that the results of resummation of the scalar
transverse energy are identical  to those of the more familiar
resummation of vector transverse momentum at order $\as$, as they
should be since at most one parton is emitted at this order.

The transverse energy computed here is the resummed component of
hadronic initial-state radiation integrated over the full range of
pseudorapidities $\eta$.  In ref.~\cite{Davies:1983di} the $E_T$ distribution
of radiation emitted in a restricted rapidity range $|\eta|< \etamax$ was
also estimated.  This was done by replacing the lower limit of
integration in Eqs.~(\ref{eq:Fcs}) by $Q_c=Q\exp(-\etamax)$, i.e.\
assuming that radiation at $q<Q_c$ does not enter the detected region.
This is justified at the leading-logarithmic level, where
$q/Q\sim\theta\sim\exp(-\eta)$ and the
scale dependence of the parton distributions and
coefficient functions in Eq.~(\ref{eq:Rcdef}) can be neglected.
Then when $\etamax=0$ the form factor $S_c$ is replaced by unity
and Eq.~(\ref{eq:resSR}) correctly predicts a delta-function at
$E_T=0$ times the Born cross section.  However, this simple
prescription cannot be correct at the NLO level, where the $\tau$
dependence of the scale must be taken into account.  Therefore we do
not consider the $E_T$ distribution in a restricted rapidity range in
the present paper. 

\subsection{Vector boson production}
One of the best studied examples of resummation is in vector boson
production through the partonic subprocess $q + \bar q'\to V$
($V=W$ or $Z$):
\beq\label{eq:DYLO}
\sigma_{c{\bar c}}^F(Q,\as(Q)) = \delta_{cq}\delta_{\bar c\bar q'}\delta(Q^2-M_V^2)\sigma_{qq'}^V\;,
\eeq
where at lowest order
\beeq
\sigma_{qq'}^W &=& \frac \pi 3\sqrt 2 G_F M_W^2 |V_{qq'}|^2\;,\nn\\
\sigma_{qq'}^Z &=& \frac \pi 3\sqrt 2 G_F M_Z^2 (V_q^2+A_q^2)\delta_{qq'}\;,
\eeeq
with $V_{qq'}$ the appropriate CKM matrix element and $V_q,A_q$ the
vector and axial couplings to the Z$^0$.
The coefficients in the quark form factor $S_q(Q,\tau)$ are~\cite{Kodaira:1982nh,Davies:1984hs}:
\beeq\label{eq:Aqetc}
&&A_q^{(1)} = C_F \;, \quad A_q^{(2)} = \frac{1}{2} C_F K \;, \quad B_q^{(1)} = - \frac{3}{2} C_F \;, \\
&&B_q^{(2)} = 
C_F^2\left(\frac{\pi^2}{4}-\frac{3}{16}-3\zeta_3\right)
+C_F\,C_A\left(\frac{11}{36}\pi^2-\frac{193}{48}+\frac{3}{2}\zeta_3\right)
+ C_F\,n_f\left(\frac{17}{24}-\frac{\pi^2}{18}\right)\nn
\eeeq
where $\zeta_n$ is the Riemann $\zeta$-function $(\zeta_3=1.202\dots)$,
$C_F=4/3$, $C_A=3$, $n_f$ is the number of light flavours, and
\beq
K = \left( \frac{67}{18} - \frac{\pi^2}{6} \right) C_A - \frac{5}{9} n_f \;\;.
\eeq

The above expression for $B_q^{(2)}$ is in a scheme where the subprocess
cross section is given by the leading-order expression (\ref{eq:DYLO}).  In the same
scheme the NLO coefficient functions are~\cite{Davies:1984hs,Balazs:1995nz}
\beeq\label{eq:CDY}
C_{qq}(\as,z)&=&\left\{1+\frac{\as}{4\pi}C_F(\pi^2-8)\right\}\delta(1-z)
+\frac{\as}{2\pi}C_F(1-z)\nn\\
&\equiv& \left(1+\frac{\as}{\pi}c^{(1)}_q\right)\delta(1-z)
+\frac{\as}{2\pi}C_F(1-z)\nn\\
C_{qg}(\as,z)&=&\frac{\as}{2\pi}z(1-z)\;,
\eeeq
where the second line defines $c_q^{(1)}$.  The corresponding splitting functions are
\beeq\label{eq:PDY}
P_{qq}(z)&=&C_F\left[\frac{1+z^2}{(1-z)_+}+\frac 32 \delta(1-z)\right] \nn\\
P_{qg}(z)&=&\frac 12\left[z^2+(1-z)^2\right]\;.
\eeeq
Equations (\ref{eq:RcNLO})--(\ref{eq:ffPplus}) therefore give
\beeq\label{eq:RqNLO}
R_q^{(R)}(\xi,\tau)&=&
\int_{\xi}^1 \frac{dx_1}{x_1}\Biggl\{f_{q/h_1}(x_1)f_{\bar q/h_2}\left(
\frac{\xi}{x_1}\right)\left(1+\frac{\as}{\pi}2c^{(1)}_q\right)\nn\\
&& + \frac{\as}{\pi}\int_{\xi/x_1}^1\frac{dz}z
\Biggl[f_{q/h_1}(x_1)f_{\bar q/h_2}\left(\frac{\xi}{zx_1}\right)C_F(1-z)\nn\\
&&+\left\{f_{g/h_1}(x_1)f_{\bar q/h_2}\left(\frac{\xi}{zx_1}\right)+ f_{q/h_1}(x_1)
f_{g/h_2}\left(\frac{\xi}{zx_1}\right)\right\}\frac 12 z(1-z)\Biggr]\Biggr\}\;,\nn\\
R_q^{(I)}(\xi,\tau)&=&
\frac{\as}2\int_{\xi}^1 \frac{dx_1}{x_1} \int_0^1\frac{dz}z
\Biggl\{2f_{q/h_1}(x_1)f_{\bar q/h_2}\left(\frac{\xi}{zx_1}\right)P_{qq}(z)\\
&&+\left[f_{g/h_1}(x_1)f_{\bar q/h_2}\left(\frac{\xi}{zx_1}\right)
+f_{q/h_1}(x_1)f_{g/h_2}\left(\frac{\xi}{zx_1}\right)\right]P_{qg}(z)\Biggr\}\nn\\
&=&\frac{\as}2\int_{\xi}^1 \frac{dx_1}{x_1}\Biggl\{
2C_F f_{q/h_1}(x_1)f_{\bar q/h_2}\left(\frac{\xi}{x_1}\right)
\left[2\ln\left(1-\frac{\xi}{x_1}\right)+\frac 32\right]\nn\\
&&+\int_{\xi/x_1}^1\frac{dz}z\Biggl[2C_F f_{q/h_1}(x_1)\left\{f_{\bar q/h_2}\left(\frac{\xi}{zx_1}\right)
\frac{1+z^2}{1-z}
- f_{\bar q/h_2}\left(\frac{\xi}{x_1}\right)\frac{2z}{1-z}\right\} \nn\\
&&+\left\{f_{g/h_1}(x_1)f_{\bar q/h_2}\left(\frac{\xi}{zx_1}\right)
+f_{q/h_1}(x_1)f_{g/h_2}\left(\frac{\xi}{zx_1}\right)\right\}
\frac 12\left\{z^2+(1-z)^2\right\}\Biggr]\Biggr\}\;. \nn
\eeeq

\subsection{Higgs boson production}
In the case of Higgs boson production the 
corresponding LO partonic subprocess is gluon
fusion, $g + g \to H$, through a massive-quark loop:
\beq\label{eq:higgsLO}
\sigma_{c{\bar c}}^F(Q,\as(Q)) = \delta_{cg}\delta_{\bar cg}\delta(Q^2-m_H^2)\sigma_0^H\;,
\eeq
where in the limit of infinite quark mass
\beq
\sigma_0^H = \frac{\as^2(m_H)G_Fm_H^2}{288\pi\sqrt{2}}\;.
\eeq
The coefficients in the gluon form factor $S_g(Q,\tau)$ are~\cite{Catani:1988vd,deFlorian:2000pr,deFlorian:2001zd}
\beeq\label{eq:Agetc}
&&A_g^{(1)} = C_A \;,\quad A_g^{(2)} = \frac{1}{2} C_A K \;,\quad
B_g^{(1)} = - \frac{1}{6} (11 C_A - 2 n_f) \;,\nn\\
&&B_g^{(2) \,H}=C_A^2\left(\frac{23}{24}+
\frac{11}{18}\pi^2-\frac{3}{2}\zeta_3\right)
+\frac{1}{2} C_F\,n_f-C_A\,n_f\left(\frac{1}{12}+\frac{\pi^2}{9} \right)
-\frac{11}{8} C_F C_A\, .
\eeeq

Here again, the above expression for $B_g^{(2)}$ is in a scheme where the Higgs subprocess
cross section is given by the leading-order expression (\ref{eq:higgsLO}).
In the same scheme the NLO coefficient functions are~\cite{Kauffman:1992cx}
\beeq\label{eq:Chiggs}
C_{gg}(\as,z)&=&\left\{1+\frac{\as}{4\pi}
\left[C_A\left(2-\frac{\pi^2}3\right)+5+4\pi^2\right]\right\}\delta(1-z)\nn\\
&\equiv& \left(1+\frac{\as}{\pi}c^{(1)}_g\right)\delta(1-z)\nn\\
C_{gq}(\as,z)&=&C_{g\bar q}(\as,z)=\frac{\as}{2\pi}C_F z\;.
\eeeq
The corresponding splitting functions are
\beeq\label{eq:Phiggs}
P_{gg}(z)&=&2C_A\left[\frac z{(1-z)_+}+\frac{1-z}z+z(1-z)\right]
+\frac 16 (11C_A-2n_f)\delta(1-z)\nn\\
P_{gq}(z)&=&P_{g\bar q}(z)= C_F\frac{1+(1-z)^2}z\;.
\eeeq
Equations (\ref{eq:RcNLO})--(\ref{eq:ffPplus}) therefore give
\beeq\label{eq:RgNLO}
&R_g^{(R)}(\xi,\tau)&=
\int_{\xi}^1 \frac{dx_1}{x_1}\Biggl\{f_{g/h_1}(x_1)f_{g/h_2}\left(
\frac{\xi}{x_1}\right)\left(1+\frac{\as}{\pi}2c^{(1)}_g\right)\nn\\
&& + \frac{\as}{\pi}\int_{\xi/x_1}^1\frac{dz}z
\left[f_{g/h_1}(x_1)f_{s/h_2}\left(\frac{\xi}{zx_1}\right)+ f_{s/h_1}(x_1)
f_{g/h_2}\left(\frac{\xi}{zx_1}\right)\right]\frac 12 C_F z\Biggr\}\;,\nn\\
&R_g^{(I)}(\xi,\tau)&=
\frac{\as}2\int_{\xi}^1 \frac{dx_1}{x_1} \int_0^1\frac{dz}z
\Biggl\{2f_{g/h_1}(x_1)f_{g/h_2}\left(\frac{\xi}{zx_1}\right)P_{gg}(z)\nn\\
&&+\left[f_{g/h_1}(x_1)f_{s/h_2}\left(\frac{\xi}{zx_1}\right)
+f_{s/h_1}(x_1)f_{g/h_2}\left(\frac{\xi}{zx_1}\right)\right]P_{gq}(z)\Biggr\}\nn\\
&=&\hspace{-0.3cm}\frac{\as}2\int_{\xi}^1 \frac{dx_1}{x_1}\Biggl\{
2f_{g/h_1}(x_1)f_{g/h_2}\left(\frac{\xi}{x_1}\right)
\left[2C_A\ln\left(1-\frac{\xi}{x_1}\right)+\frac 16 (11C_A-2n_f)\right]\nn\\
&&\hspace{-1cm}+\int_{\xi/x_1}^1\hspace{-0.2cm}\frac{dz}z\Biggl[4C_Af_{g/h_1}(x_1)\left\{f_{g/h_2}\left(\frac{\xi}{zx_1}\right)
\left[\frac{z}{1-z}+\frac{1-z}z+z(1-z)\right]
- f_{g/h_2}\left(\frac{\xi}{x_1}\right)\frac{z}{1-z}\right\} \nn\\
&&\hspace{-1cm}+\left\{f_{g/h_1}(x_1)f_{s/h_2}\left(\frac{\xi}{zx_1}\right)
+f_{s/h_1}(x_1)f_{g/h_2}\left(\frac{\xi}{zx_1}\right)\right\}C_F\frac{1+(1-z)^2}z\Biggr]\Biggr\} 
\eeeq
where $f_s = \sum_q(f_q+f_{\bar q})$.

\section{Resummed distributions}\label{sec:resum}
\subsection{Vector boson production}

\begin{figure}
\begin{center}
\epsfig{file=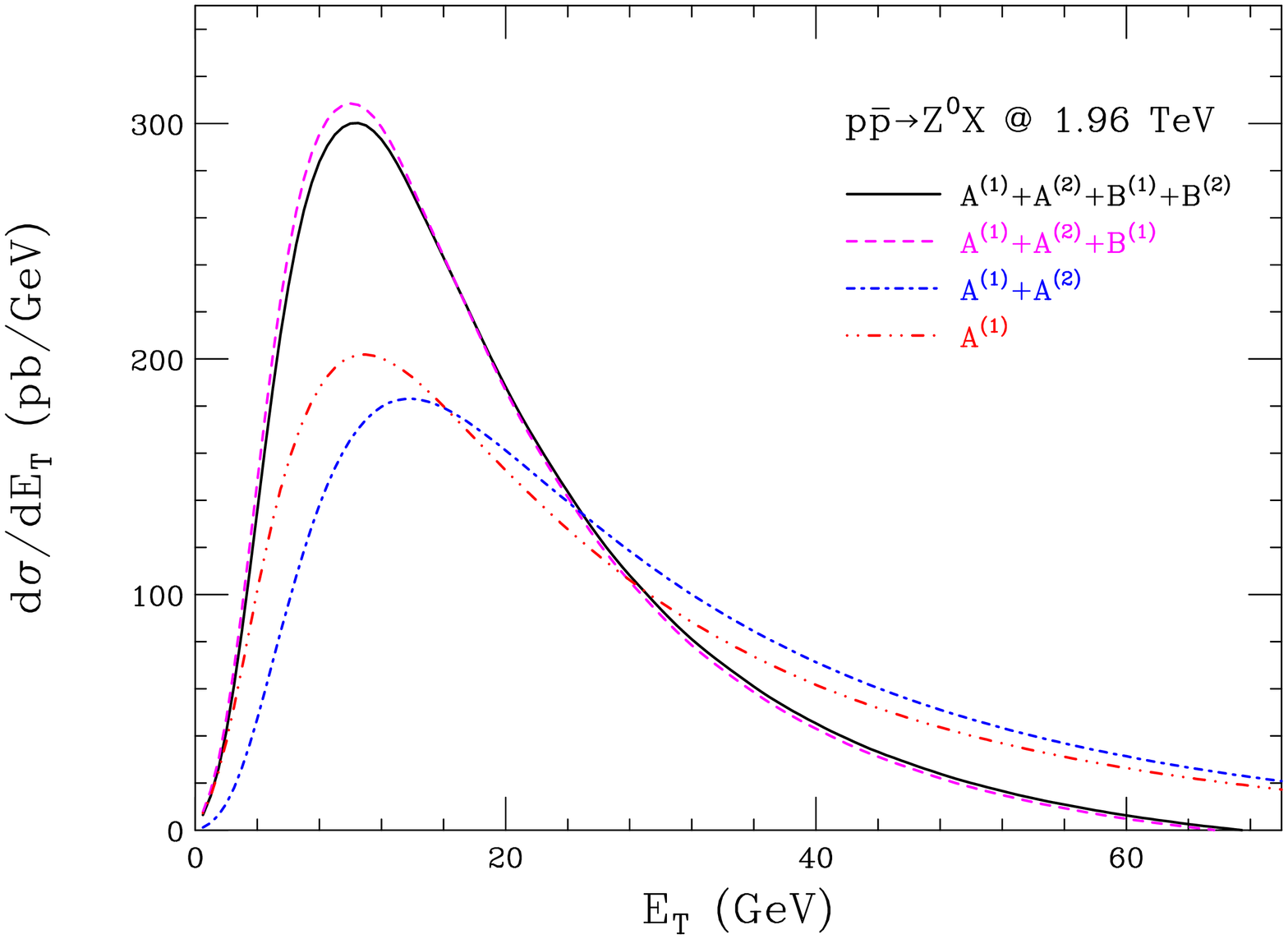,angle=90,width=75mm}
\epsfig{file=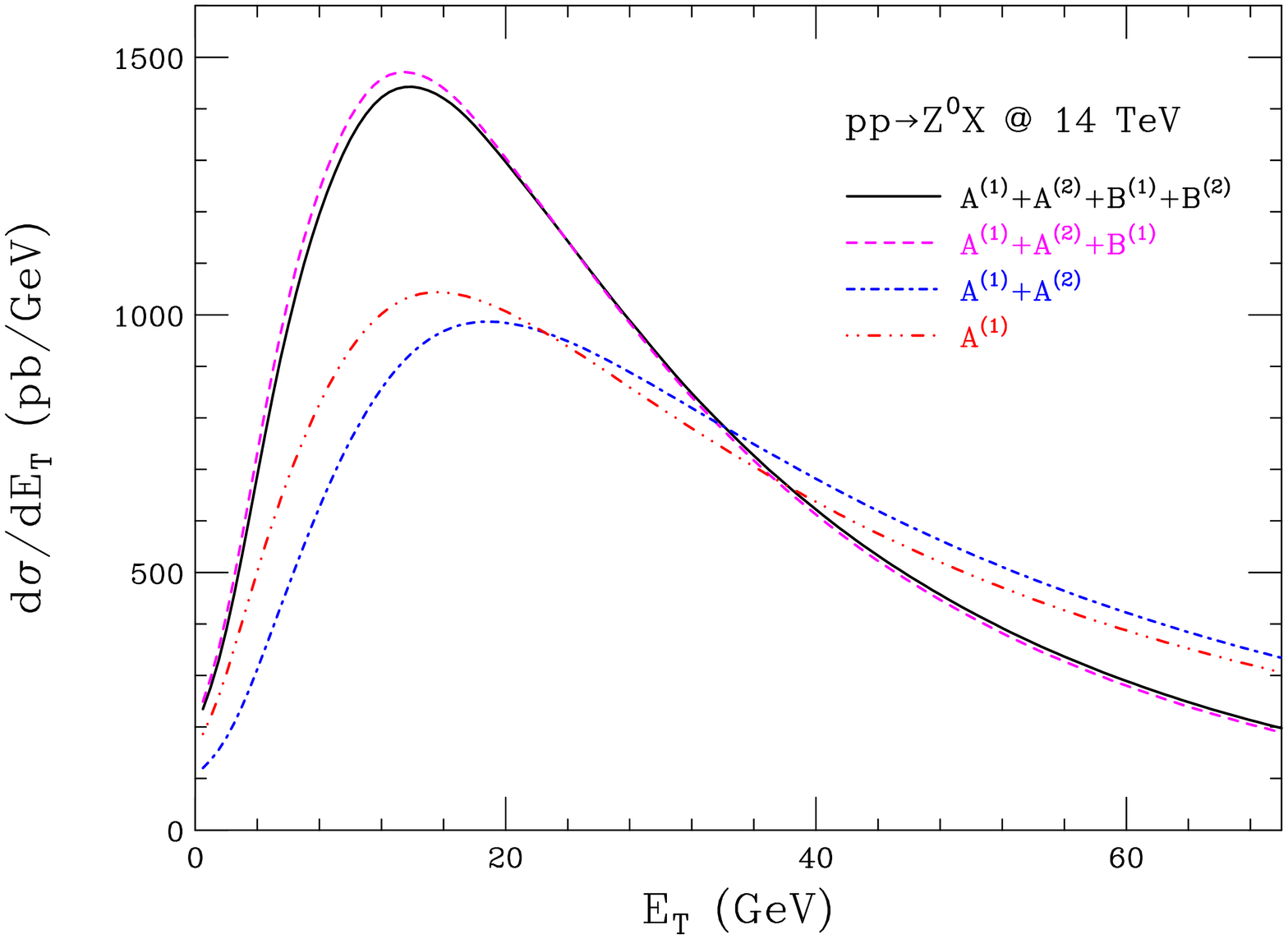,angle=90,width=75mm}
\end{center}
\caption{Resummed component of the transverse energy distribution in Z$^0$ boson production at the Tevatron
and LHC.  The curves show the effects of the coefficients in the quark form factor:
black, all coefficients; magenta omitting $B_q^{(2)}$; blue $A_q^{(1)}$ and $A_q^{(2)}$ only; red $A_q^{(1)}$ only.
\label{fig:ETdrelyTev} }
\end{figure}

Figure~\ref{fig:ETdrelyTev} shows the resummed component of the
transverse energy distribution in Z$^0$ boson production at the
Tevatron  ($p\bar p$ at $\rs =1.96$ TeV) and LHC ($pp$ at $\rs
=14$ TeV).\footnote{Results for $pp$ at $\rs =7$ TeV are given in
  Appendix B.}  For all calculations, we use the MSTW 2008 NLO parton
distributions~\cite{Martin:2009iq}.  The different curves show the
effects of the subleading coefficients (\ref{eq:Aqetc}) in the quark
form factor.  We see that while $B_q^{(1)}$ has a large effect (the
difference between the blue and magenta curves), the effects of the
other subleading coefficients are quite small.

The peak of the resummed distribution lies at around
$E_T\sim 10$ GeV at the Tevatron, rising to $\sim 14$ GeV at the LHC.
This is comfortably below $M_Z$, justifying the resummation of
logarithms of $E_T/M_Z$ in the peak region.  However, at LHC energy
the predicted distribution has a substantial tail at larger values of $E_T$,
indicating that the higher-order terms generated by the resummation
formula remain significant even when the logarithms are not large.
In addition, the LHC prediction does not go to zero as it should at
small $E_T$.  However, this region is sensitive to the treatment of
non-perturbative effects such as the behaviour of the strong coupling
at low scales (we freeze its value below 1 GeV) and the upper limit in
the integral over transverse time (we set $\tau_{\mbox{\scriptsize
  max}}=1/\Lambda$ where $\Lambda$ is the two-loop
QCD scale parameter, set to 200 MeV here).

The resummed component for W$^\pm$ boson production looks very
similar, apart of course from the overall normalization, and therefore
we do not show it here.  Predictions with matching to fixed order will
be presented in Section~\ref{sec:match}.

\subsection{Higgs boson production}

\begin{figure}
\begin{center}
\epsfig{file=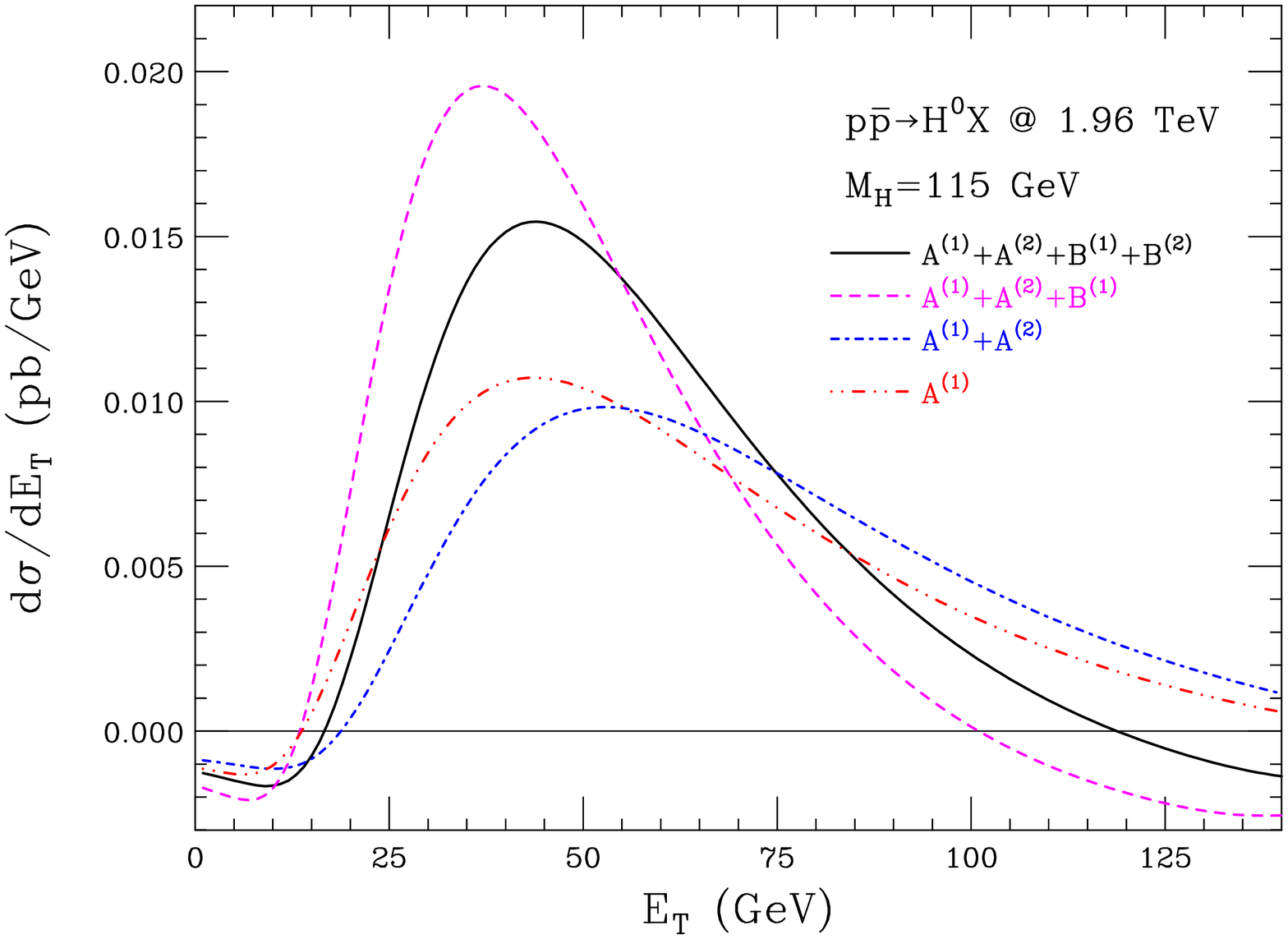,angle=90,width=75mm}
\epsfig{file=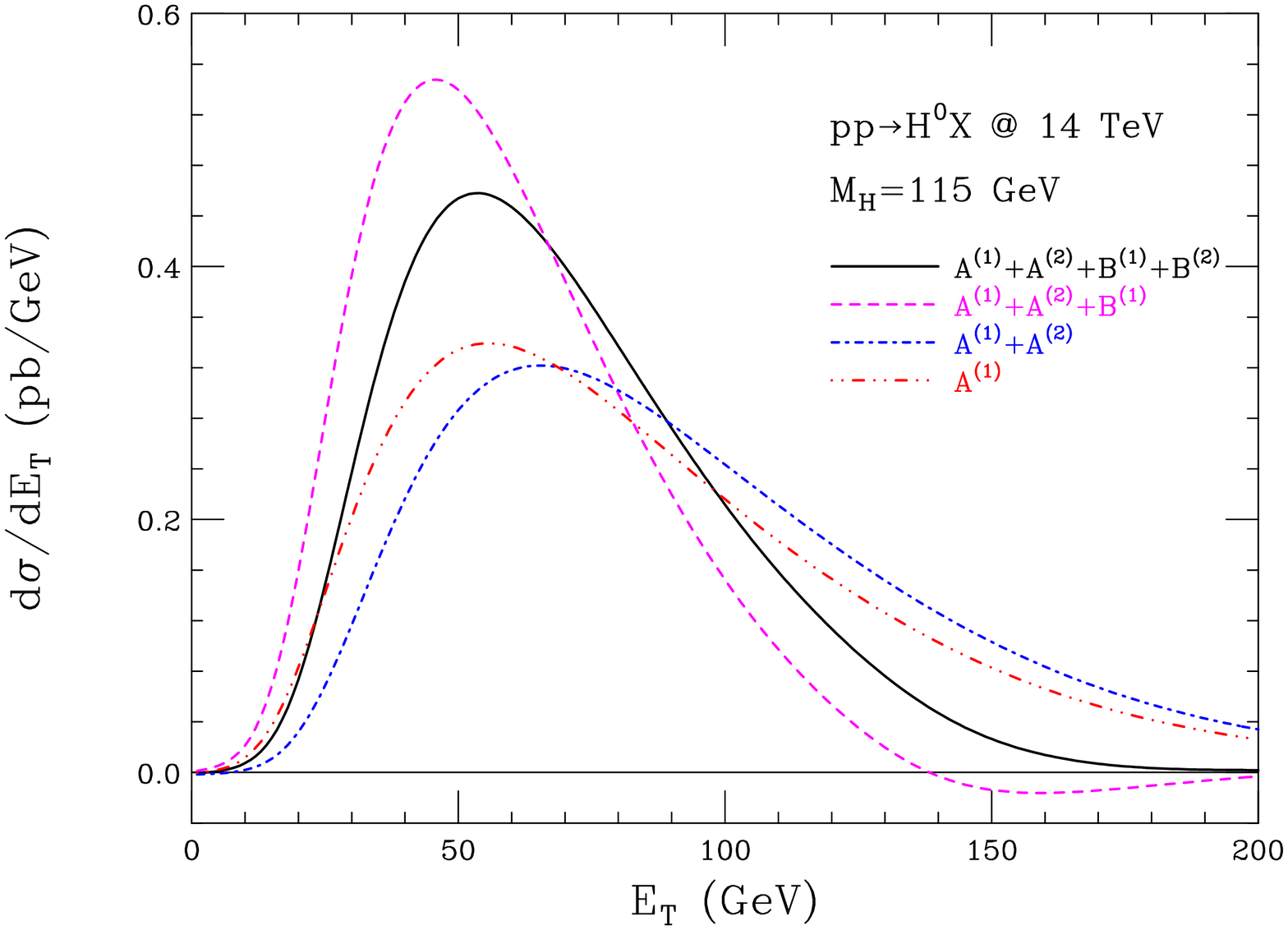,angle=90,width=75mm}
\end{center}
\caption{Resummed component of the transverse energy distribution in
  Higgs boson production at the Tevatron and LHC. The curves shows
the effects of the coefficients in the gluon form factor: black, all
coefficients; magenta omitting $B_g^{(2)}$; blue $A_g^{(1)}$
and $A_g^{(2)}$ only; red $A_g^{(1)}$ only.
\label{fig:ETh115Tev} }
\end{figure}

Figure~\ref{fig:ETh115Tev} shows the resummed component of the
transverse energy distribution in Higgs boson production at the
Tevatron and LHC, for a Higgs mass of 115 GeV.  The effects
of subleading terms in the gluon form factor (\ref{eq:Agetc}) are more
marked than those of the quark form factor discussed above.  The
distribution peaks at large values of $E_T$, around 40 GeV at
the Tevatron, rising to $\sim 50$ GeV at the LHC.  This is due to the
larger colour charge of the gluon.  However, together with the large
effects of subleading terms, it does make the reliability of the
resummed predictions more questionable.  Also in contrast to
the vector boson case, the suppression at low and high $E_T$ is if
anything too great, resulting in negative values below 16 GeV and
above 120 GeV at Tevatron energy.

\section{Matching to fixed order}\label{sec:match}

The resummed distributions presented above include only terms that are
logarithmically enhanced at small $E_T$.  To extend the predictions to
larger $E_T$ we must match the resummation to fixed-order
calculations.  To avoid double counting of the resummed terms, the
corresponding contribution must be subtracted from the fixed-order
result.

We consider here only matching to first order in $\as$.  To this order
the $E_T$ distribution for $E_T>0$ has the form
\beq\label{eq:Oas}
\frac{d\sigma}{dE_T} = \frac 1{E_T}(A\ln E_T +B) + C(E_T)
\eeq
where $A$ and $B$ are constants (for a given process and collision
energy) and the function $C(E_T)$ is regular at $E_T=0$.  The terms
involving $A$ and $B$ are already included in the resummed prediction,
and therefore we have only to add the regular function $C$ to it to obtain a
prediction that is matched to the ${\cal O}(\as)$ result.  This
function is determined by fitting the ${\cal O}(\as)$ prediction
for $E_T\,d\sigma/dE_T$ to a linear function of $\ln E_T$ at small
$E_T$, extracting the coefficients $A$ and $B$, and then subtracting the
enhanced terms in Eq.~(\ref{eq:Oas}).

\subsection{Vector boson production}

\begin{figure}
\begin{center}
\epsfig{file=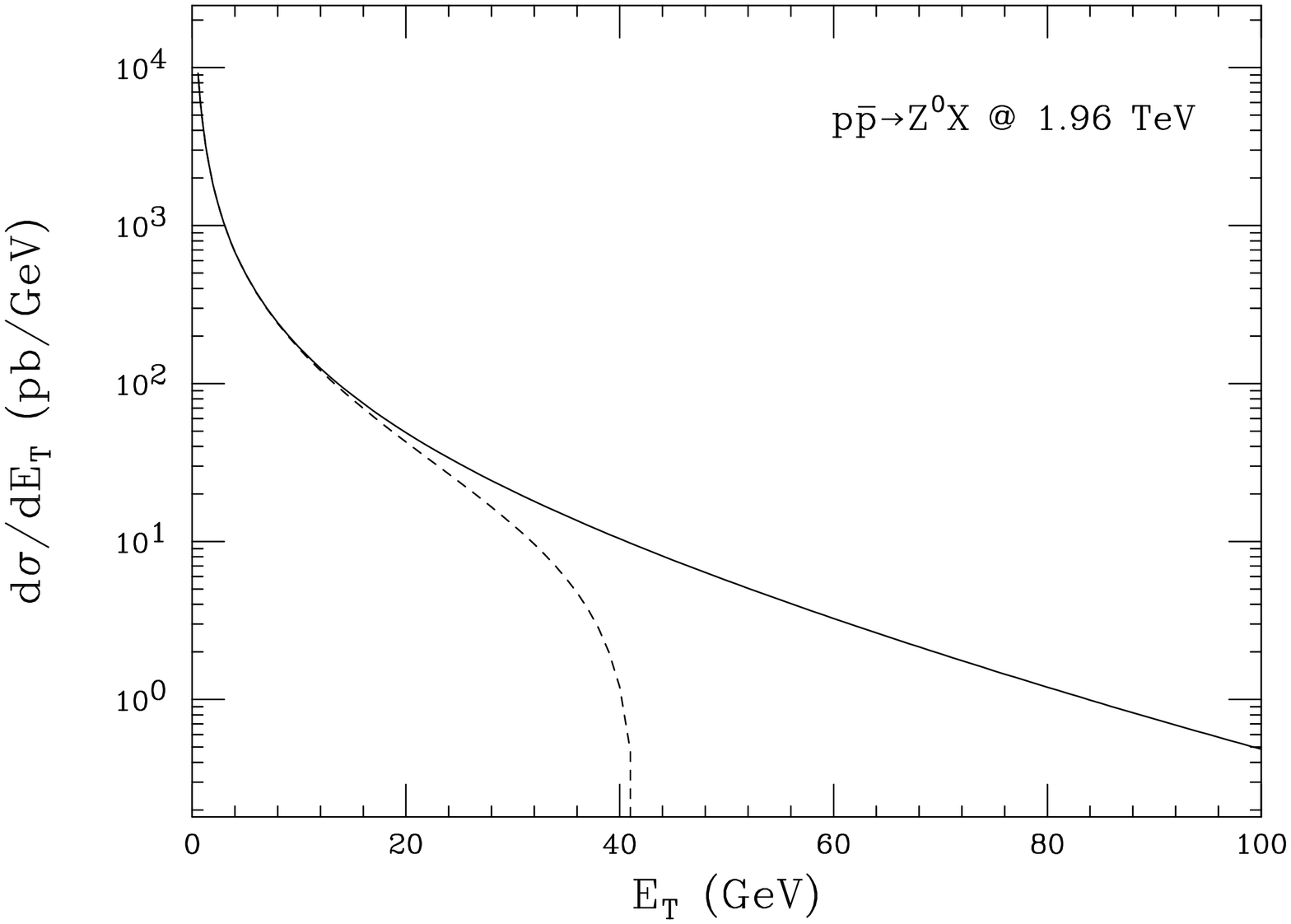,angle=90,width=75mm}
\epsfig{file=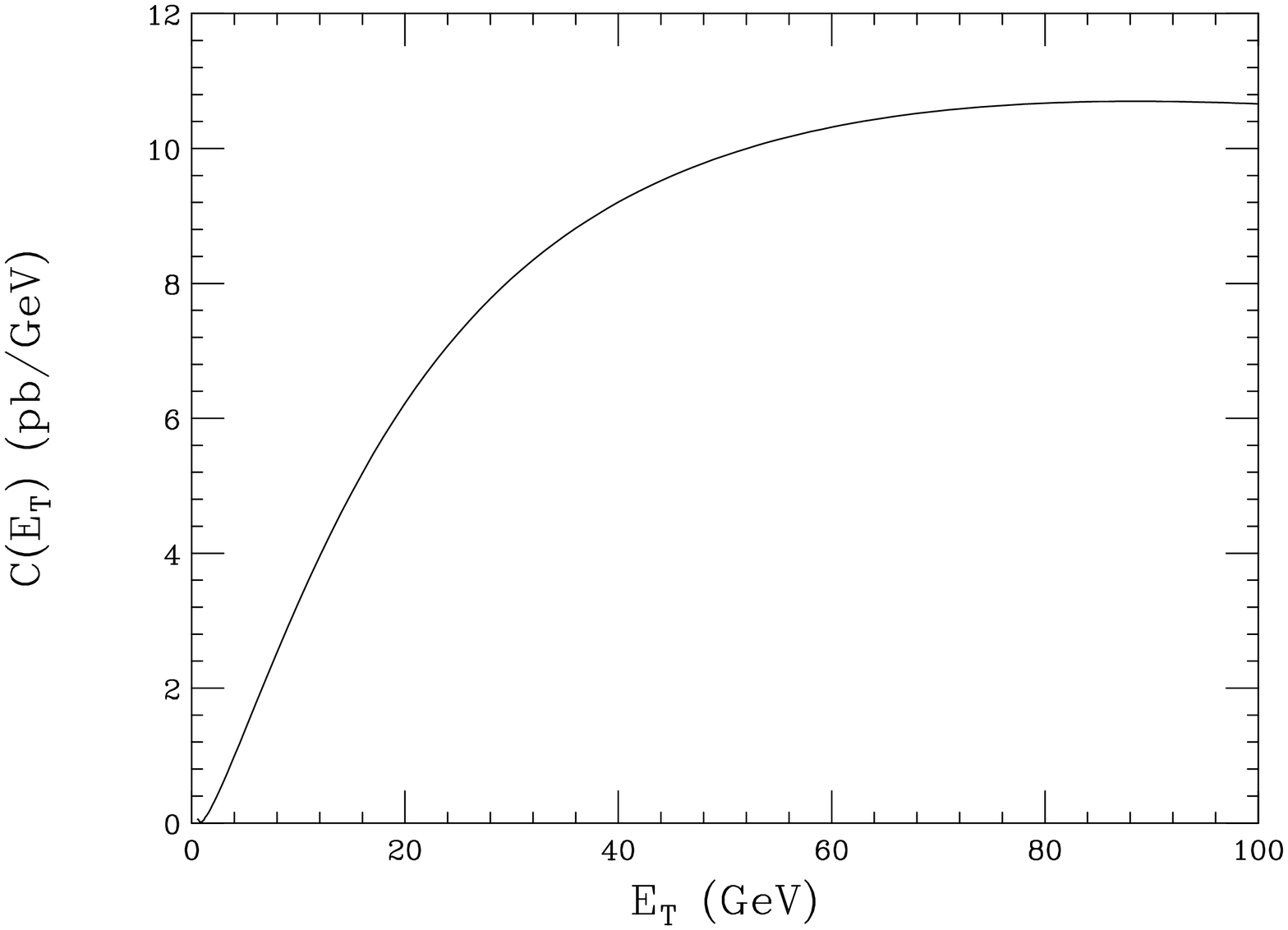,angle=90,width=75mm}
\end{center}
\caption{Left: order-$\as$ $E_T$ distribution in Z$^0$ production at the Tevatron;
solid, full prediction; dashed, fit to enhanced terms.
Right: difference between full prediction and fit to enhanced terms.
\label{fig:ZfitTev1} }
\end{figure}

\begin{figure}
\begin{center}
\epsfig{file=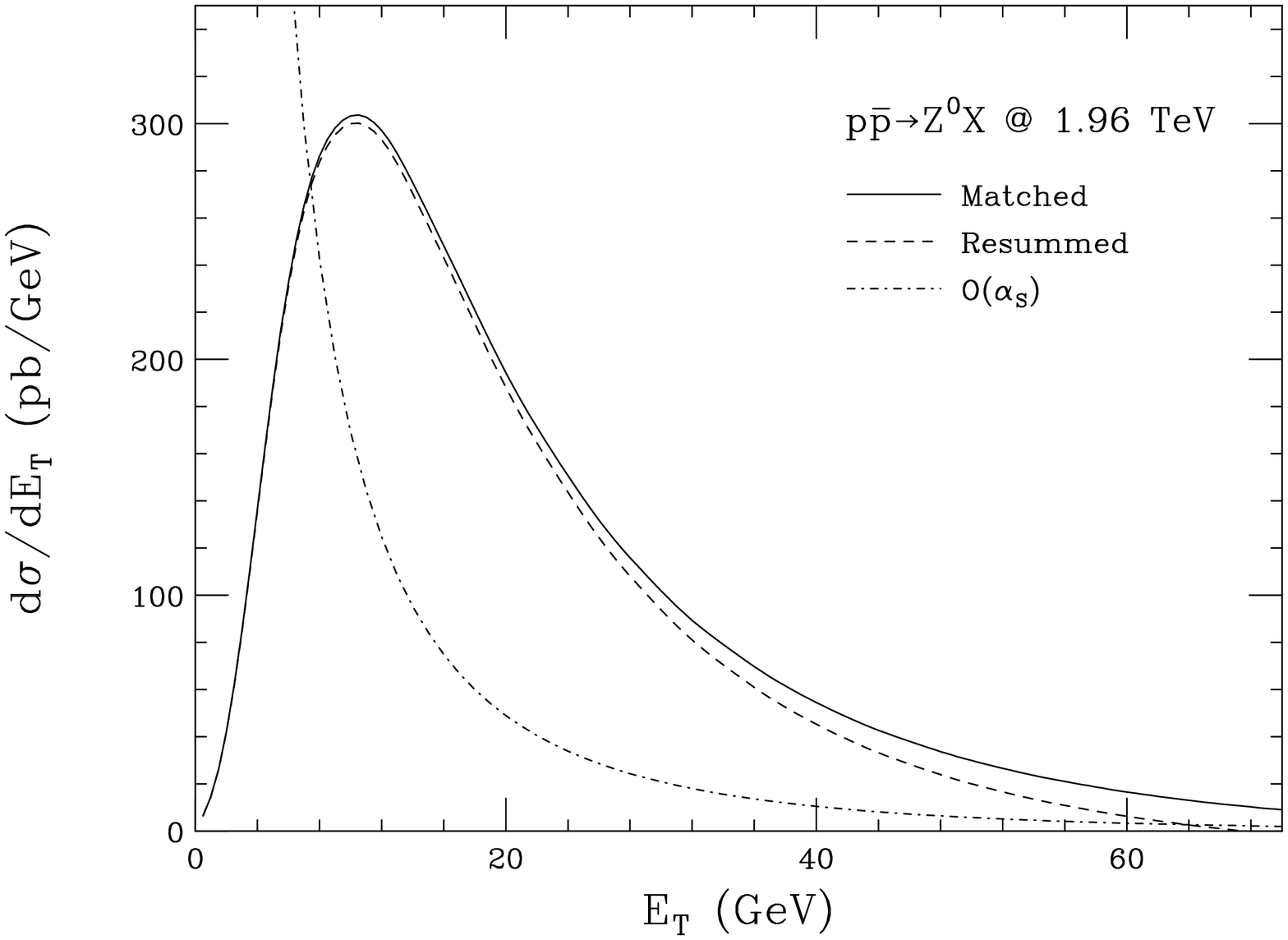,angle=90,width=75mm}
\epsfig{file=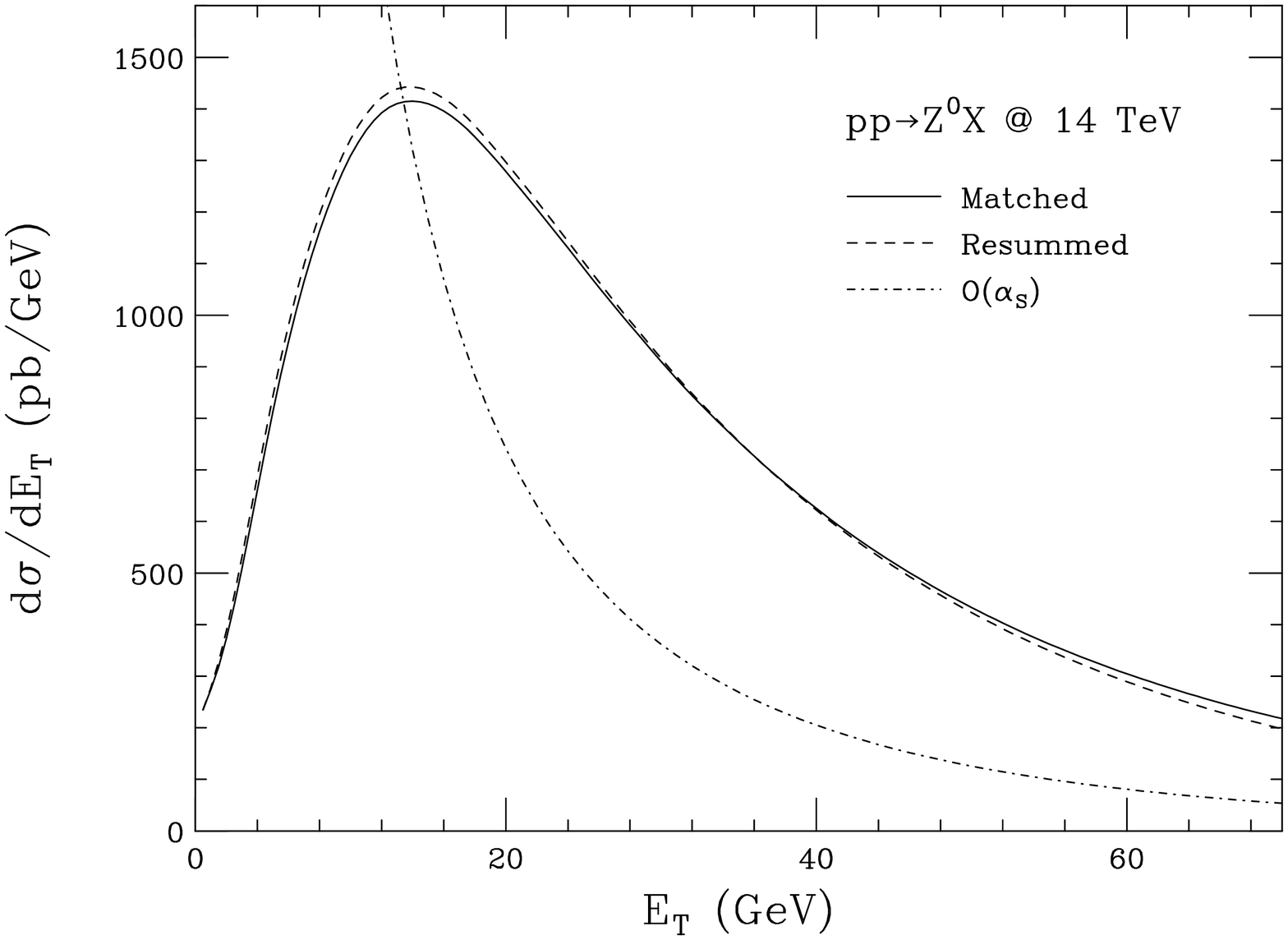,angle=90,width=75mm}
\end{center}
\caption{Predicted $E_T$ distribution in Z$^0$ production  at the
  Tevatron and LHC.  Solid: resummed prediction matched to ${\cal
    O}(\as)$.  Dashed: resummed only. Dot-dashed: ${\cal O}(\as)$ only.
\label{fig:ZfitTev3} }
\end{figure}

The above matching procedure is illustrated for  Z$^0$ production at
the Tevatron in Fig.~\ref{fig:ZfitTev1}.  The fit to the
logarithmically enhanced terms gives excellent agreement with the
order-$\as$ result out to around 20 GeV, confirming the dominance of
such terms throughout the region of the peak in
Fig.~\ref{fig:ETdrelyTev}.  The remainder function $C(E_T)$ vanishes
at small $E_T$ and rises to around 10 pb/GeV, falling off slowly at
large $E_T$. Consequently the matching correction to the resummed
prediction is small and roughly constant throughout the region 40--100
GeV, as shown in Fig.~\ref{fig:ZfitTev3}.

As shown on the right in Fig.~\ref{fig:ZfitTev3}, the situation is
similar at LHC energy: the matching correction is small, although in
this case it is negative below about 40 GeV.  The large tail at high
$E_T$ and the bad behaviour at low $E_T$, due to uncompensated
higher-order terms generated by resummation, are not much affected by
matching to this order.

\begin{figure}
\begin{center}
\epsfig{file=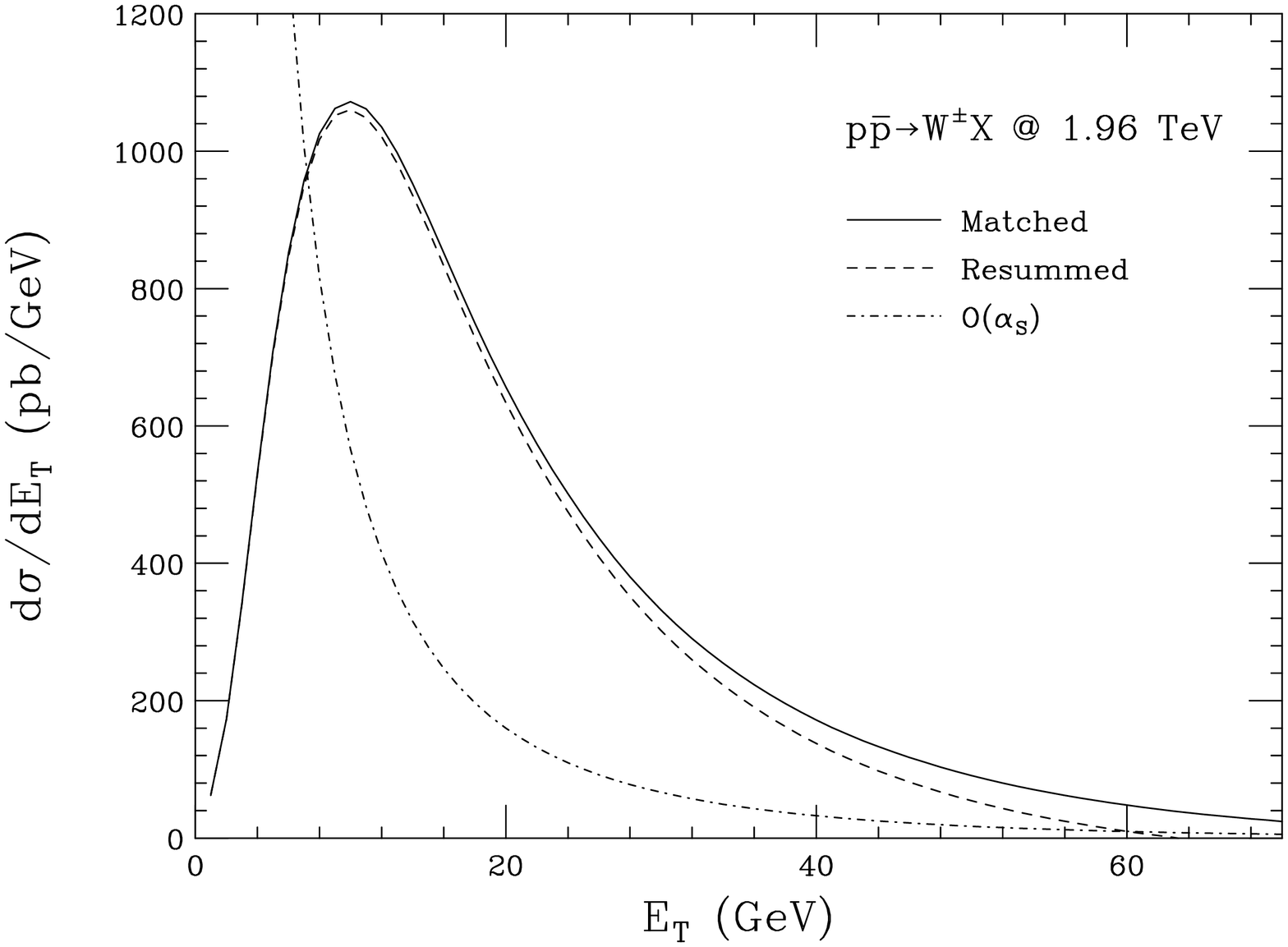,angle=90,width=75mm}
\epsfig{file=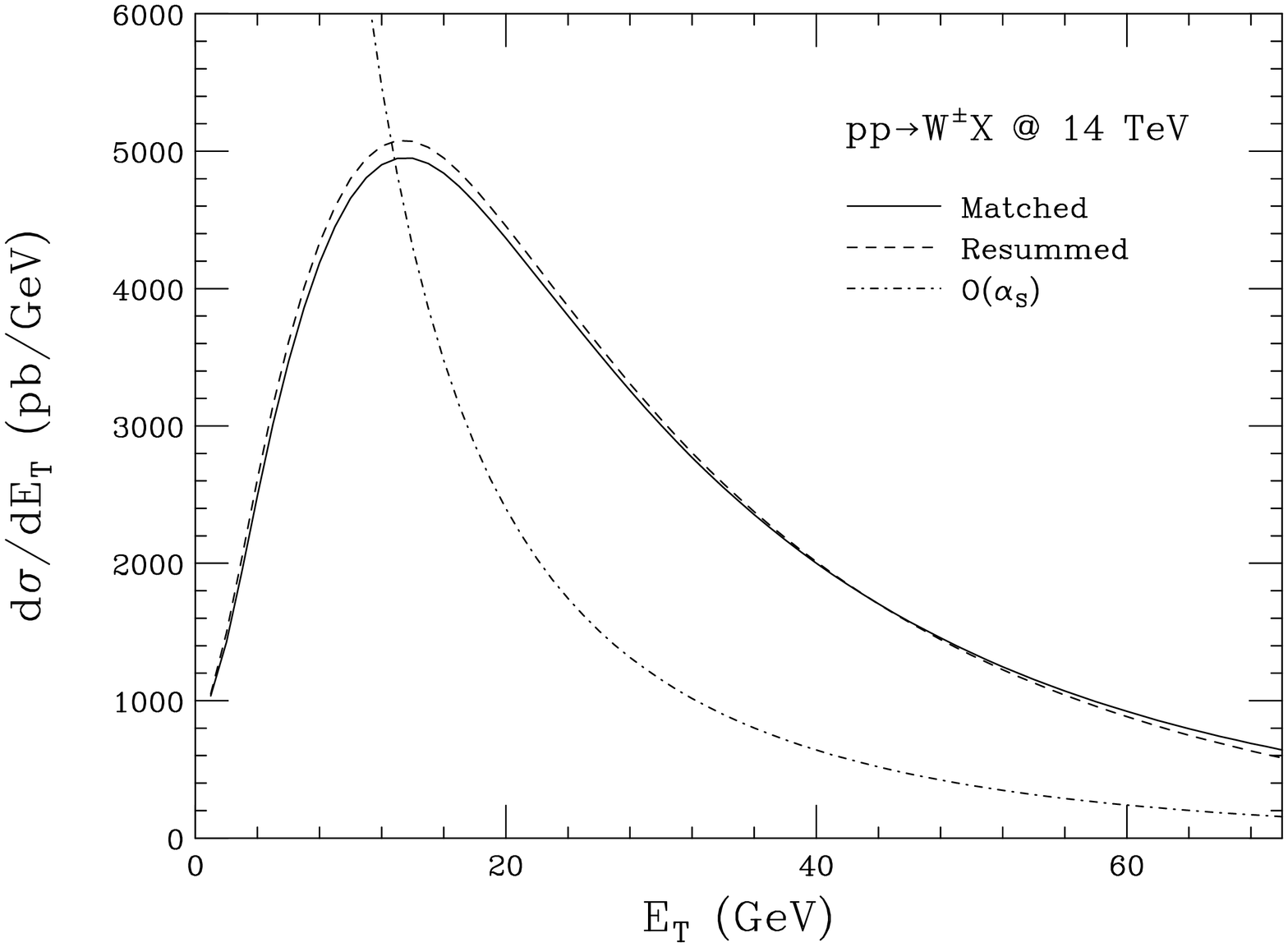,angle=90,width=75mm}
\end{center}
\caption{Predicted $E_T$ distribution in W$^+$+W$^-$ production  at the
  Tevatron and LHC.  Solid: resummed prediction matched to ${\cal
    O}(\as)$.  Dashed: resummed only. Dot-dashed: ${\cal O}(\as)$ only.
\label{fig:WfitLHC3} }
\end{figure}

The corresponding matched predictions for W$^\pm$ boson production are shown
in Fig.~\ref{fig:WfitLHC3}.  As remarked earlier, the form of the
resummed distribution is very similar to that for Z$^0$ boson production,
and again the matching correction is small.

\subsection{Higgs boson production}

\begin{figure}
\begin{center}
\epsfig{file=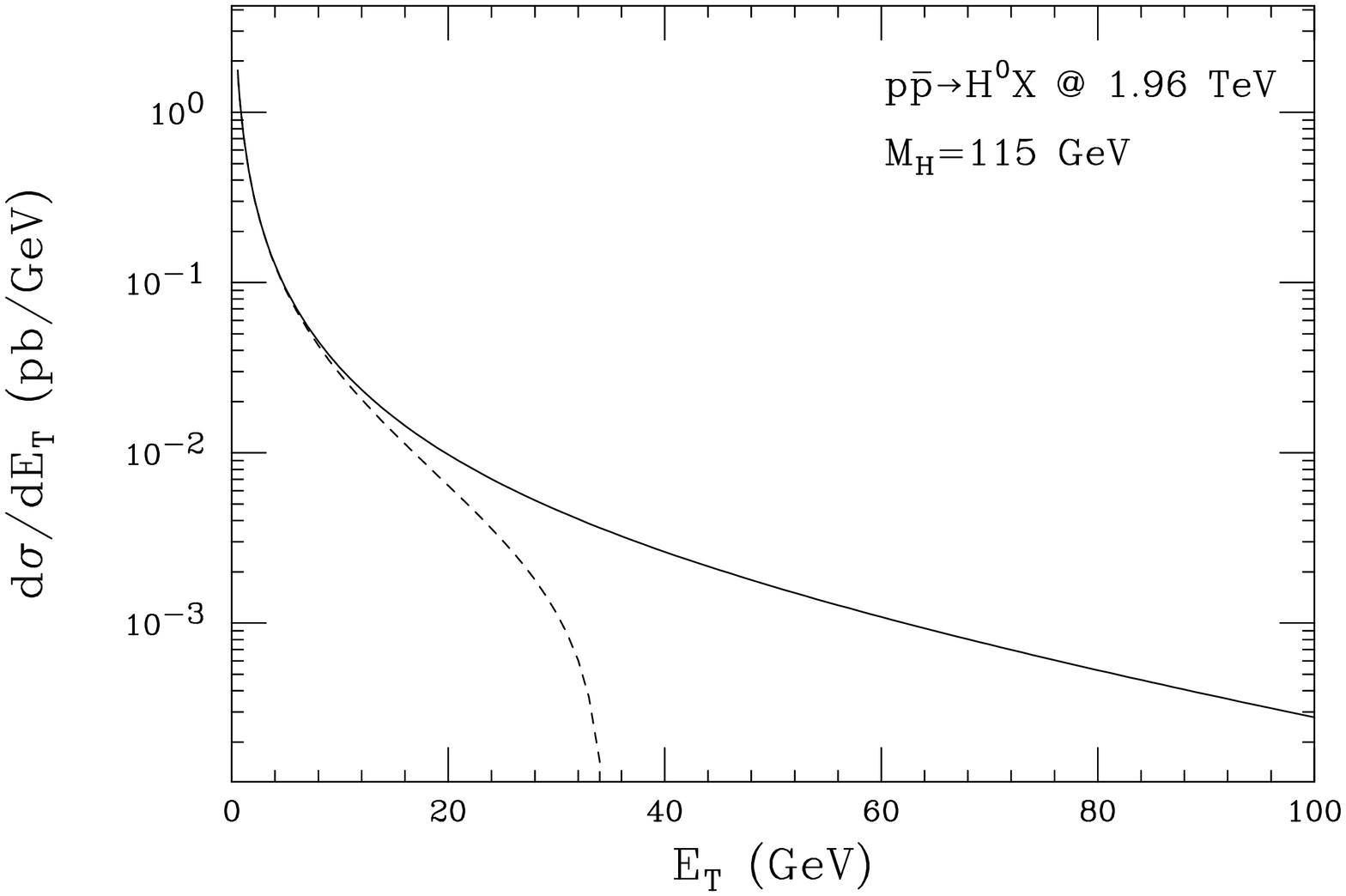,angle=90,width=75mm}
\epsfig{file=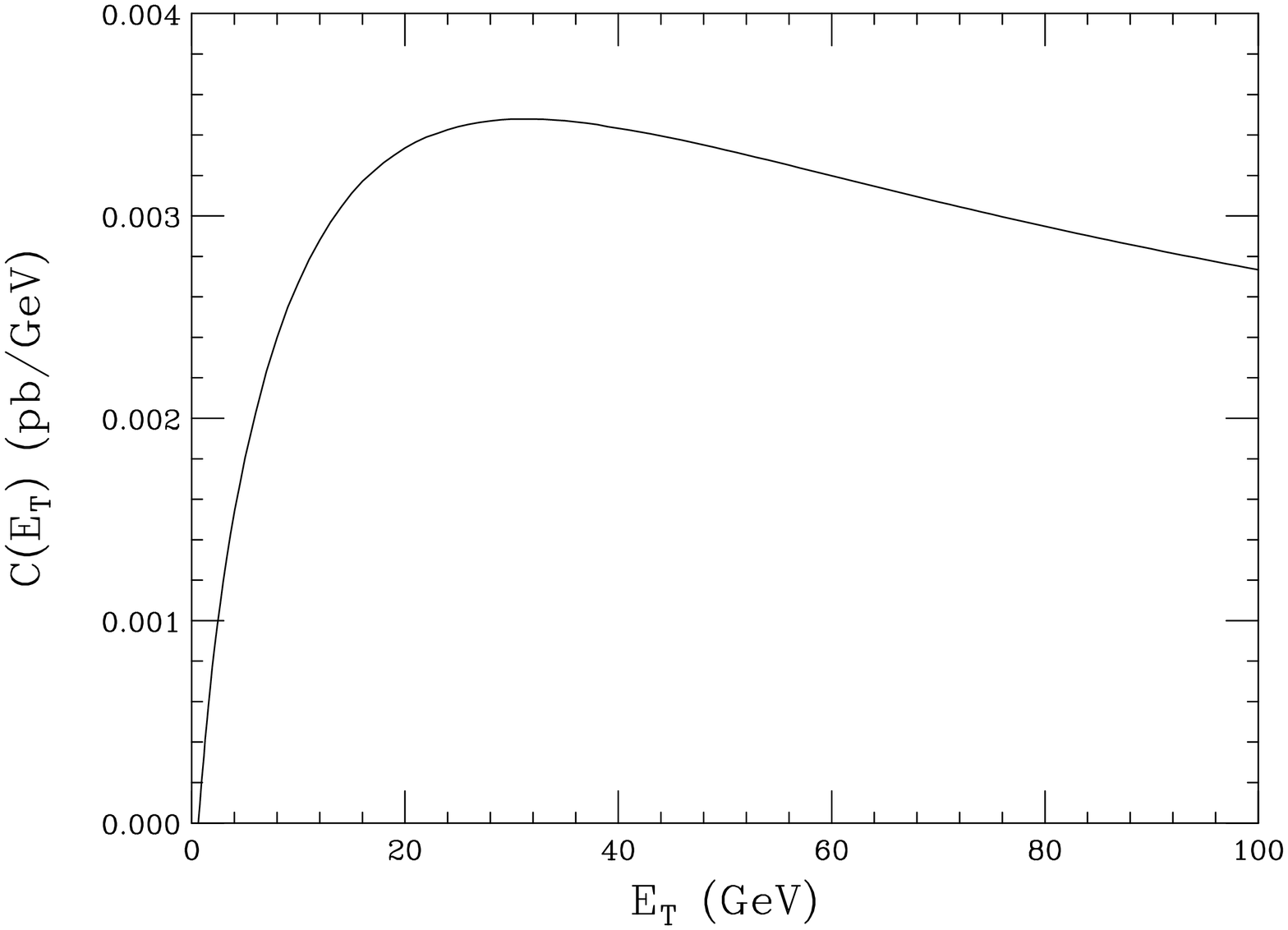,angle=90,width=75mm}
\end{center}
\caption{ Left: order-$\as$ $E_T$ distribution in Higgs boson
  production at the Tevatron;  solid, full prediction; dashed, fit to enhanced terms.
Right: difference between full prediction and fit to enhanced terms.
\label{fig:HfitTev1} }
\end{figure}

\begin{figure}
\begin{center}
\epsfig{file=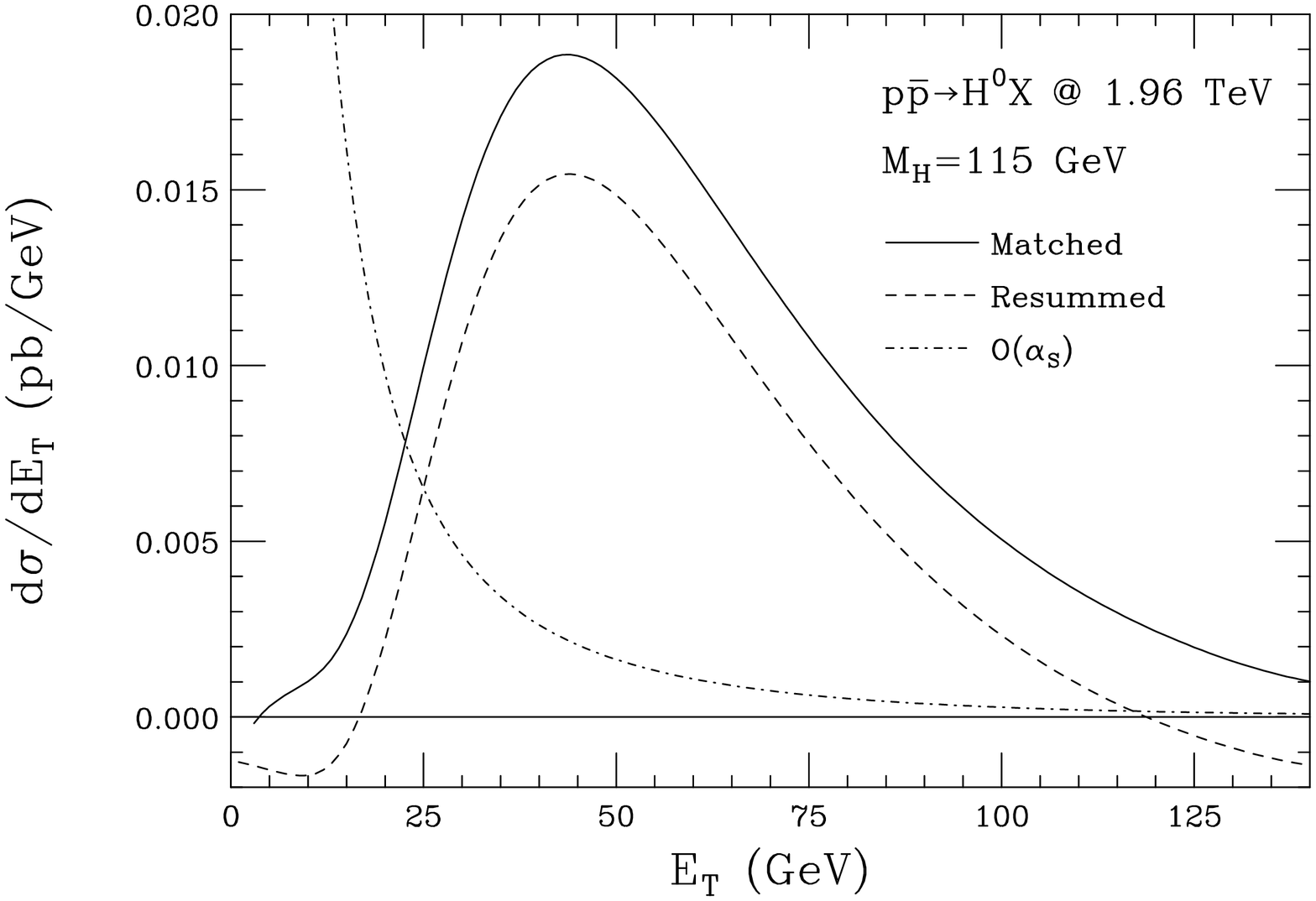,angle=90,width=75mm}
\epsfig{file=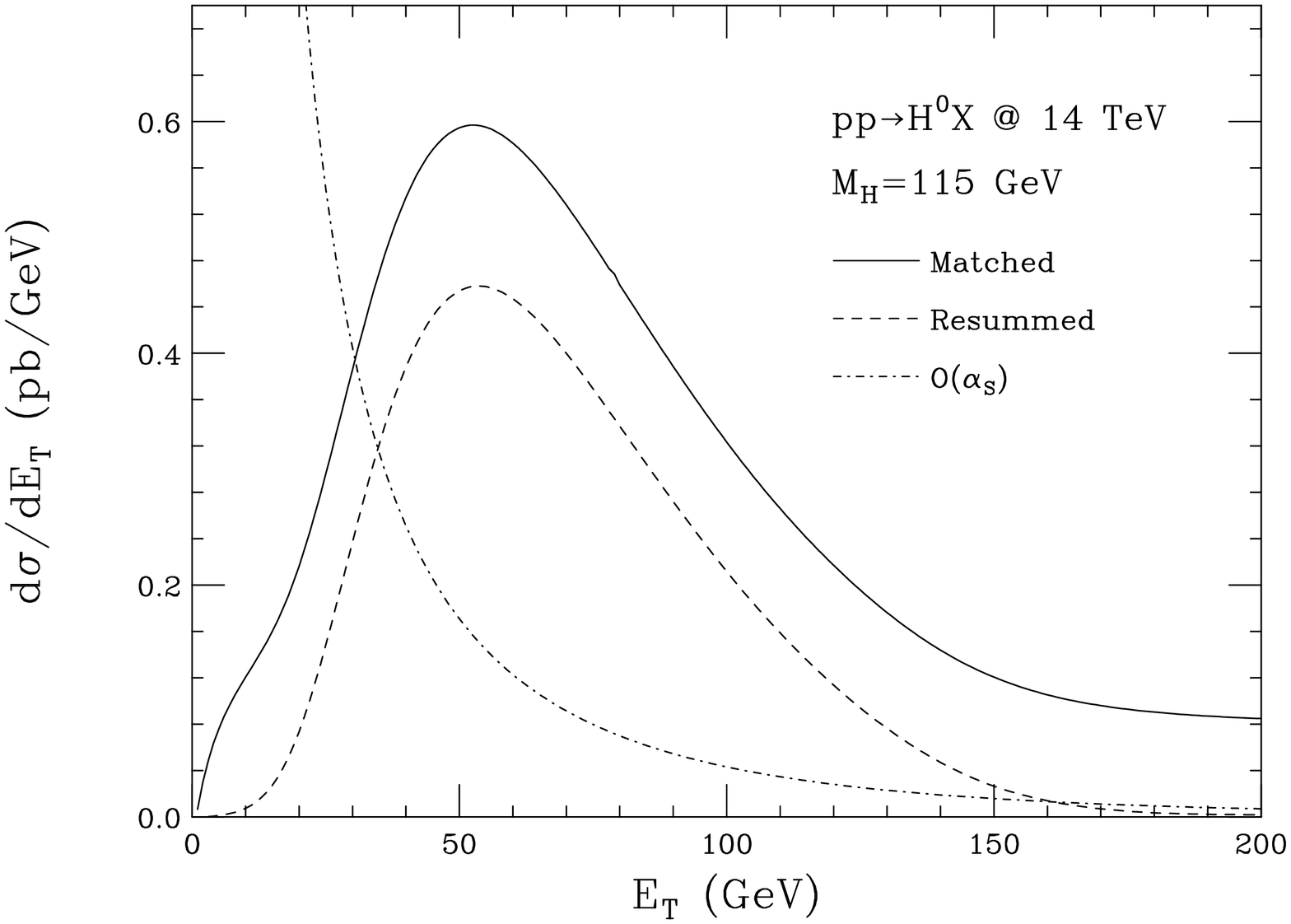,angle=90,width=75mm}
\end{center}
\caption{Predicted $E_T$ distribution in Higgs boson production at the
  Tevatron and LHC.  Solid: resummed prediction matched to ${\cal
    O}(\as)$.  Dashed: resummed only. Dot-dashed: ${\cal O}(\as)$ only.
\label{fig:HfitTev3} }
\end{figure}

Adopting the same matching procedure for Higgs boson production, we
find the results shown in Figs.~\ref{fig:HfitTev1} and
\ref{fig:HfitTev3}.  The form of the matching correction is similar to
that for vector bosons, but its effect is rather different. The
roughly constant, then slowly decreasing, correction in the region 20--100 GeV is
not small compared to the resummed result and therefore it raises the whole
distribution by a significant amount throughout this region.
This has the beneficial effect of compensating the negative values
at low and high $E_T$ at Tevatron energy.  However, it further
enhances the high-$E_T$ tail of the distribution at LHC energy.  This,
together with the relatively large correction in the peak region,
casts further doubt on the reliability of the predictions in the case
of Higgs production.

\section{Monte Carlo comparisons}\label{sec:MC}

In this section we compare the resummed and matched distributions
obtained above with the predictions of the parton shower Monte Carlo
programs {\tt HERWIG} \cite{Corcella:2000bw} and {\tt Herwig++} \cite{Bahr:2008pv}.

Comparisons are performed first at the parton level, that is, after QCD showering
from the incoming and outgoing partons of the hard subprocess.
We say ``incoming and outgoing'' because both programs apply hard matrix
element corrections: in addition to the Born process, order-$\as$ real
emission hard subprocesses are included in phase-space regions not covered
by showering from the Born process.

After showering, the Monte Carlo programs apply a hadronization model
to convert the partonic final state to a hadronic one.  We show the
effects of hadronization in the case of  {\tt HERWIG} only; those in
{\tt Herwig++} are broadly similar since both programs use basically
the same cluster hadronization model.  The programs also model the
underlying event, which arises from the interactions of spectator
partons and makes a significant contribution to the hadronic
transverse energy.  In this case we show only the underlying event
prediction of  {\tt Herwig++}, since the default model used in {\tt HERWIG}
has been found to give an unsatisfactory description of Tevatron data.
For an improved simulation of the underlying event, {\tt HERWIG} can
be interfaced to the multiple interaction package
{\tt JIMMY}~\cite{Butterworth:1996zw}, which
is similar to the model built into {\tt Herwig++}.

\subsection{Vector boson production}

\begin{figure}
\begin{center}
\epsfig{file=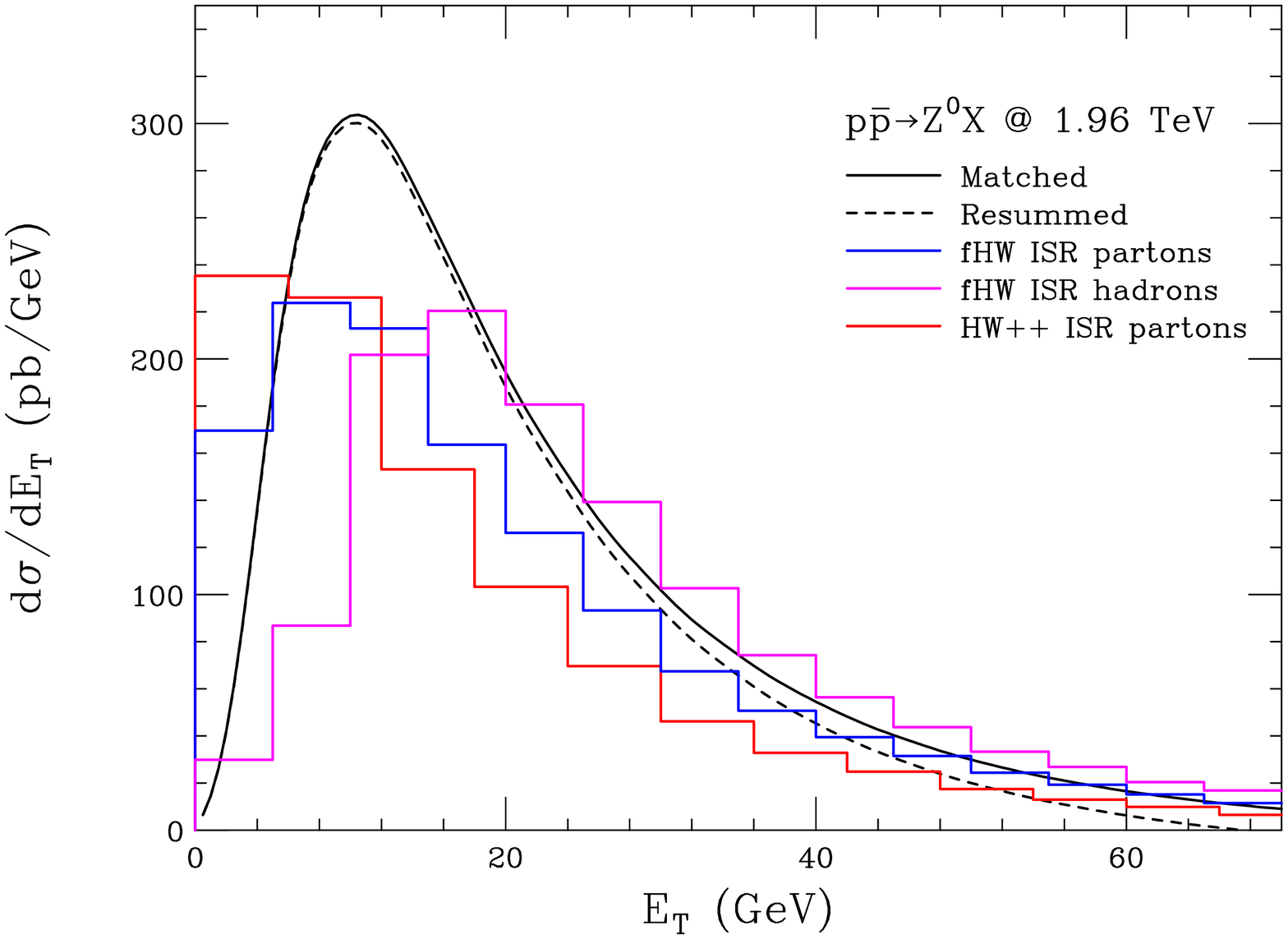,angle=90,width=75mm}
\epsfig{file=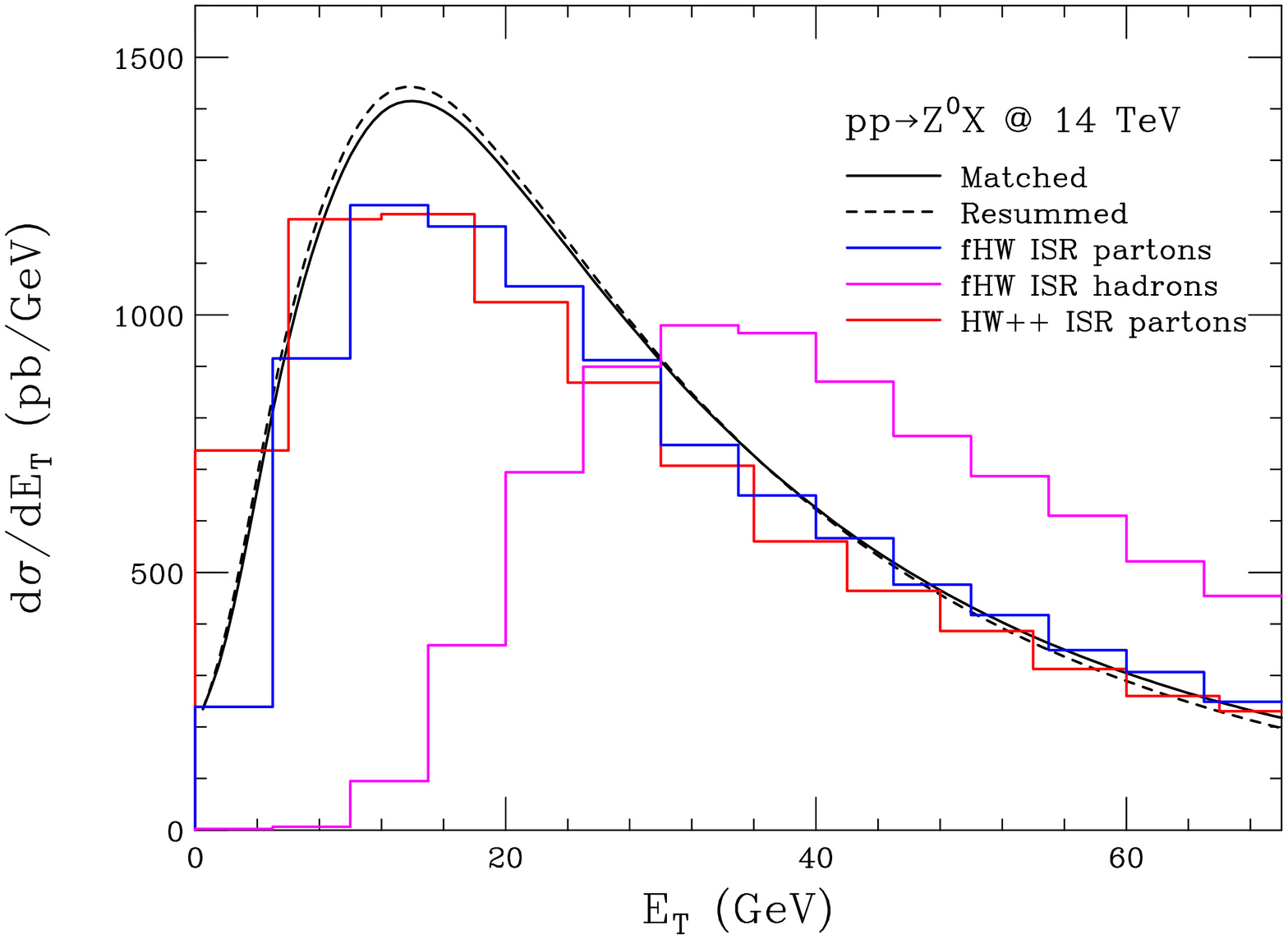,angle=90,width=75mm}
\end{center}
\caption{Predicted $E_T$ distribution in Z$^0$ boson production  at the
 Tevatron and LHC. Comparison of resummed and Monte Carlo results.
\label{fig:ZfHWall} }
\end{figure}

\begin{figure}
\begin{center}
\epsfig{file=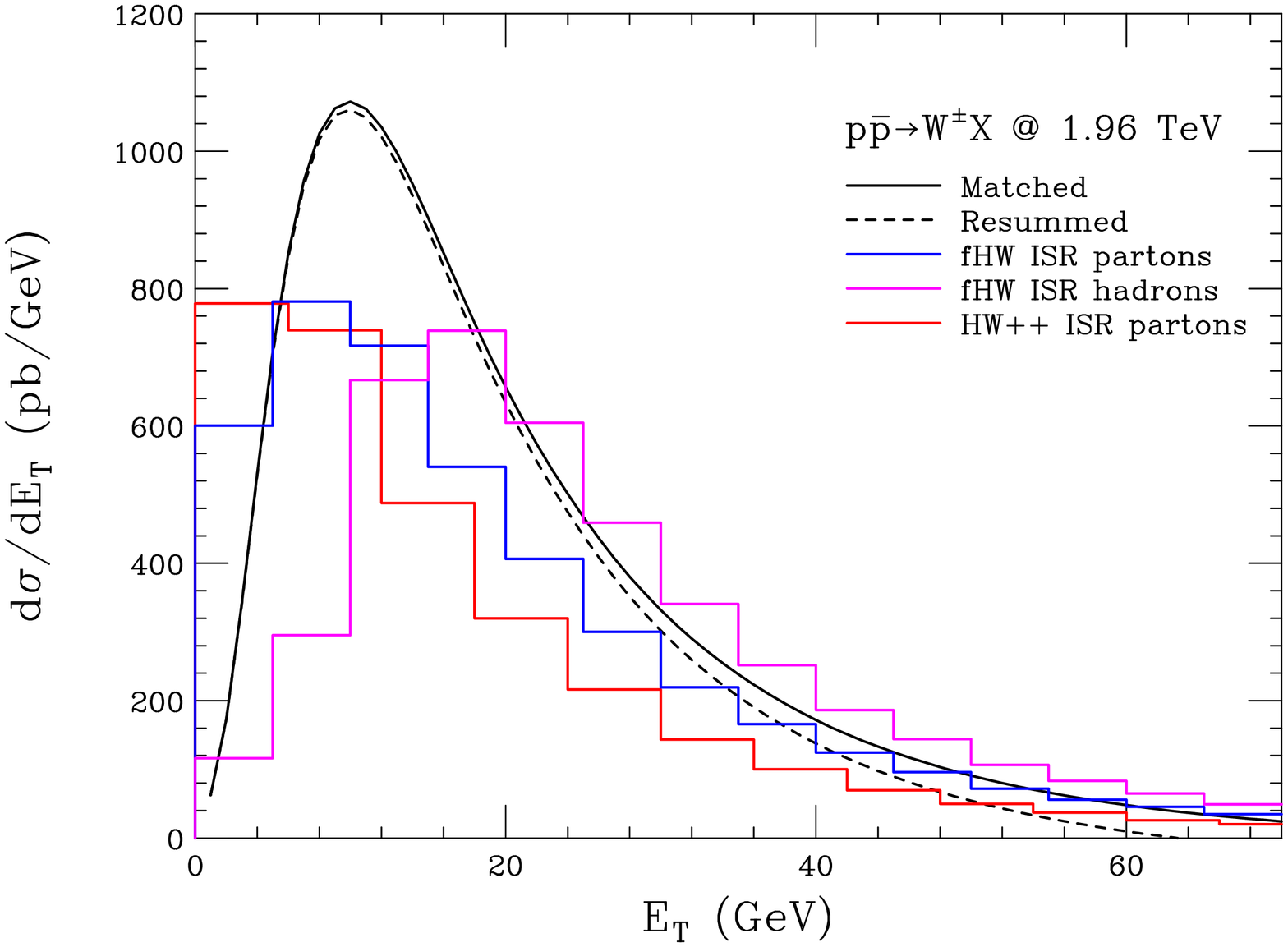,angle=90,width=75mm}
\epsfig{file=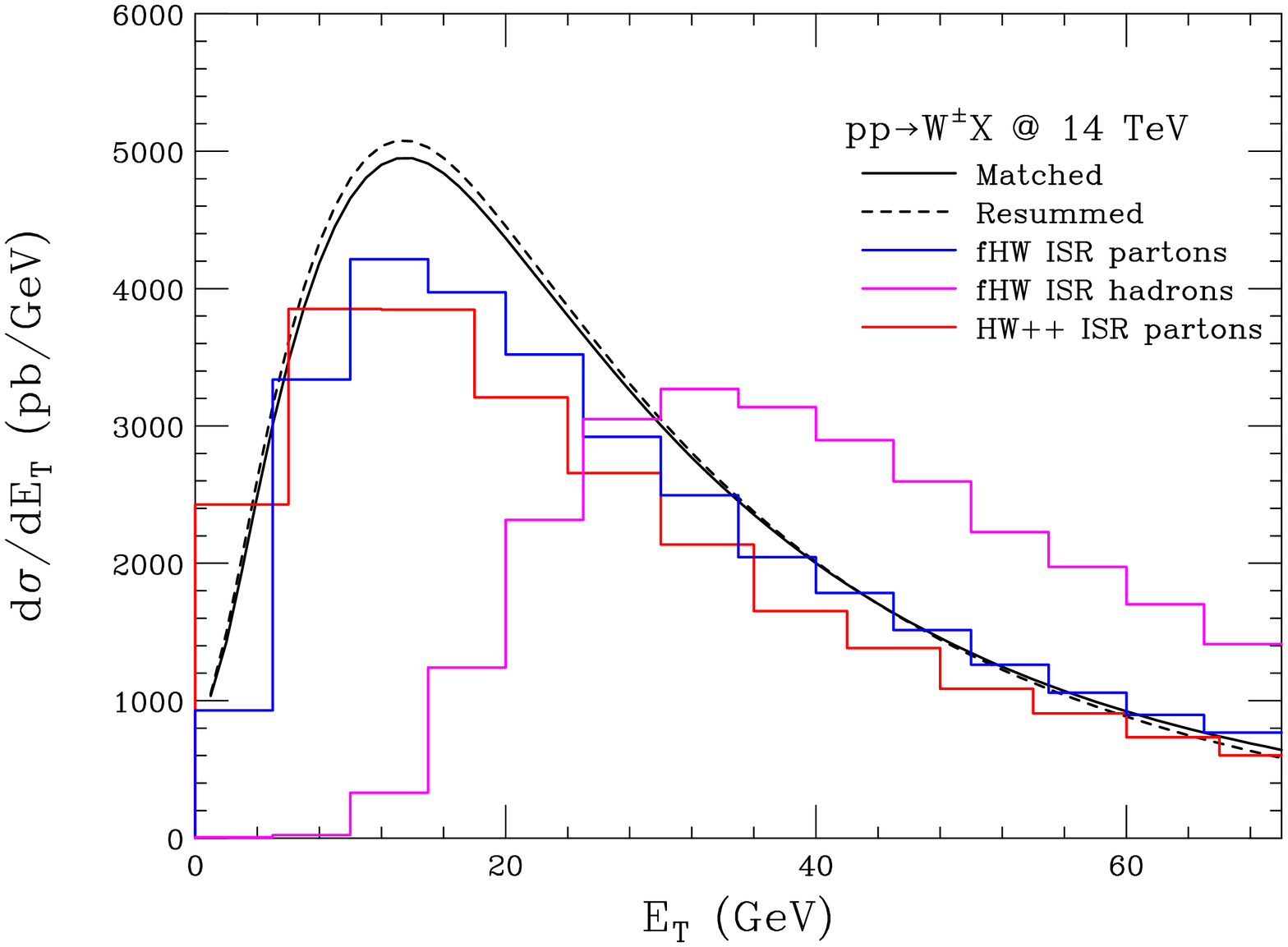,angle=90,width=75mm}
\end{center}
\caption{Predicted $E_T$ distribution in W$^+$+W$^-$ boson production  at the
 Tevatron and LHC. Comparison of resummed and Monte Carlo results.
\label{fig:WfHWall} }
\end{figure}

Figure~\ref{fig:ZfHWall} shows the comparisons for Z$^0$ production at
the Tevatron and LHC.  The {\tt HERWIG} predictions are renormalized by
a factor of 1.3 to account for the increase in the cross section from
LO to NLO.  The  {\tt Herwig++} results were not renormalized, because they
were obtained using LO** parton distributions \cite{Sherstnev:2007nd}, 
which aim to reproduce the NLO cross section.  We see that the parton-level 
Monte Carlo predictions of both programs agree fairly well with the matched
resummed results above about 15 GeV, but  {\tt Herwig++} generates a
substantially higher number of events with low values of $E_T$.  A
similar pattern is evident in the results on W$^\pm$ boson production,
shown in Fig.~\ref{fig:WfHWall}.  The effects of hadronization, shown by the
difference between the blue and magenta histograms, are also similar for
both vector bosons.  They generate a significant shift in the
distribution, of around 10 GeV at Tevatron energy and 20 GeV at LHC.

\subsection{Higgs boson production}

\begin{figure}
\begin{center}
\epsfig{file=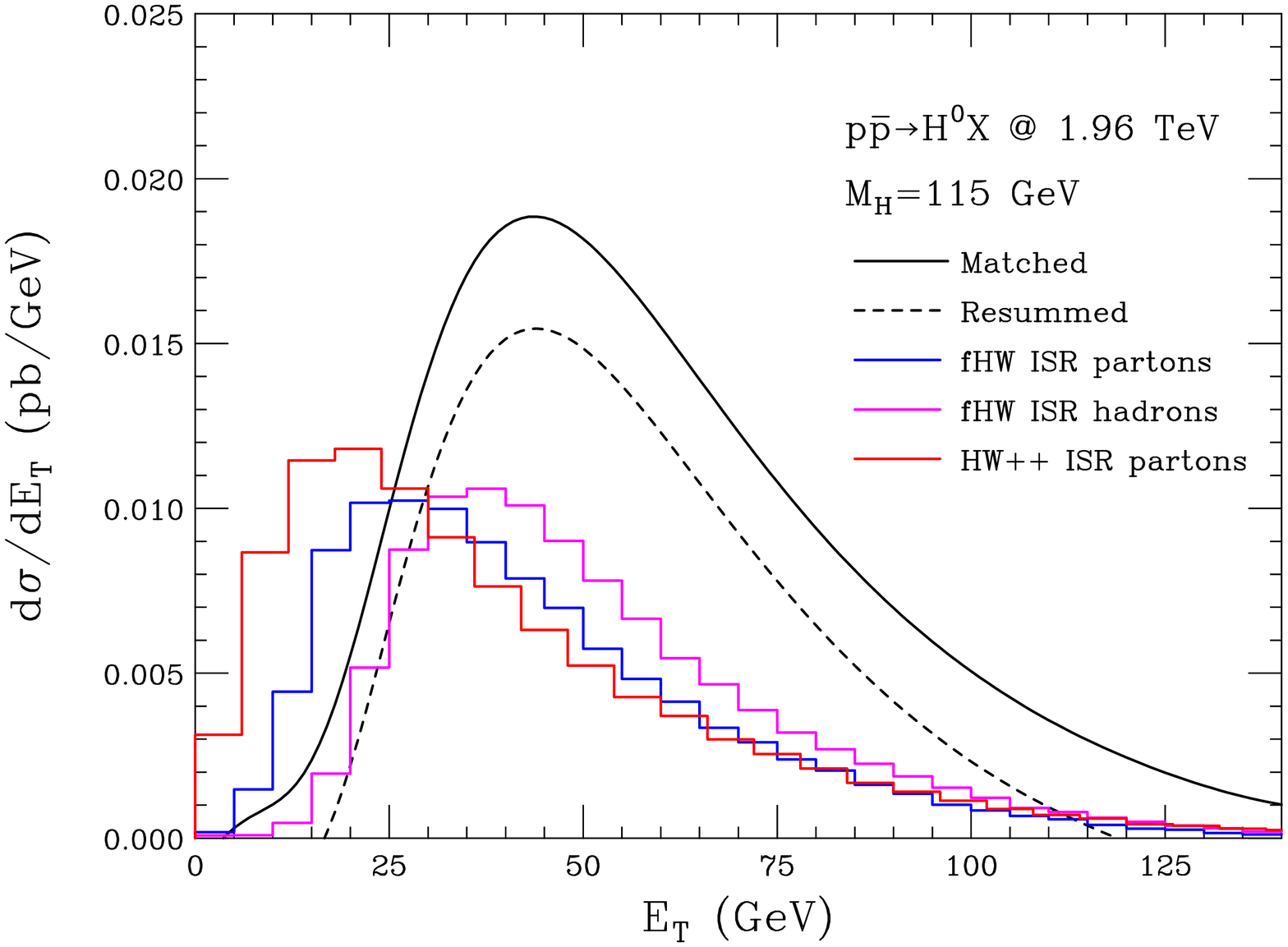,angle=90,width=75mm}
\epsfig{file=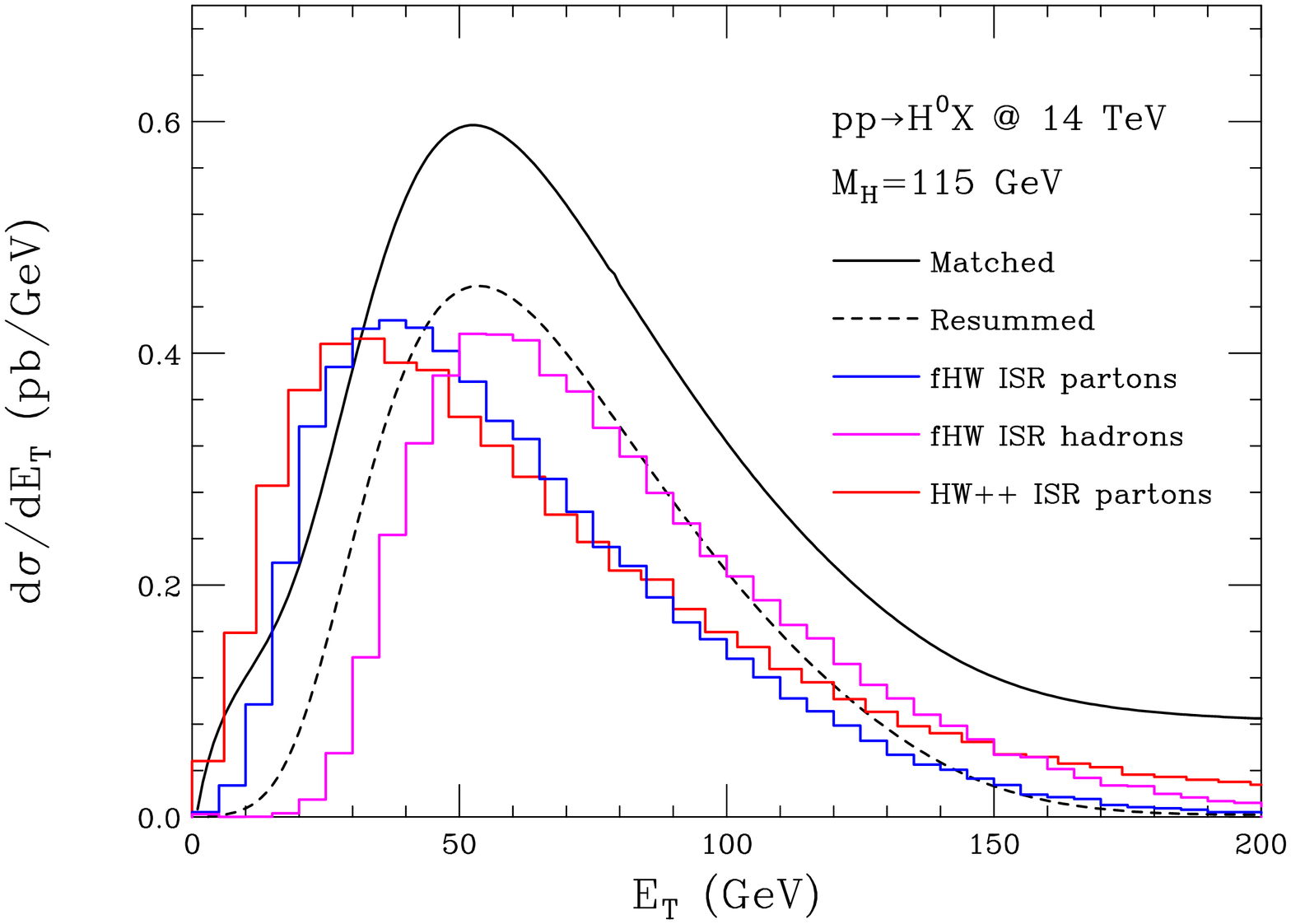,angle=90,width=75mm}
\end{center}
\caption{Predicted $E_T$ distribution in Higgs boson production  at the
 Tevatron and LHC. Comparison of resummed and Monte Carlo results.
\label{fig:HfHWall} }
\end{figure}

As may be seen from Fig.~\ref{fig:HfHWall}, the agreement between the
resummed and parton-level Monte Carlo results is less good in the case
of Higgs boson production than it was for vector bosons.  Here we have
renormalized the  {\tt HERWIG} predictions by a factor of 2 to allow for
the larger NLO correction to the cross section.  Then the Monte Carlo $E_T$
distributions agree quite well with each other but fall well below the
matched resummed predictions.  Fair agreement above about 40 GeV can
be achieved by adjusting the normalization, but then the Monte Carlos
predict more events at lower $E_T$. The effect of
 hadronization is similar to that in vector boson production, viz.\ a
 shift of about 10 GeV at the Tevatron rising to 20 GeV at the LHC,
 which actually brings the {\tt HERWIG} distribution into
 somewhat better agreement with the resummed result.

\subsection{Modelling the underlying event}

\begin{figure}
\begin{center}
\epsfig{file=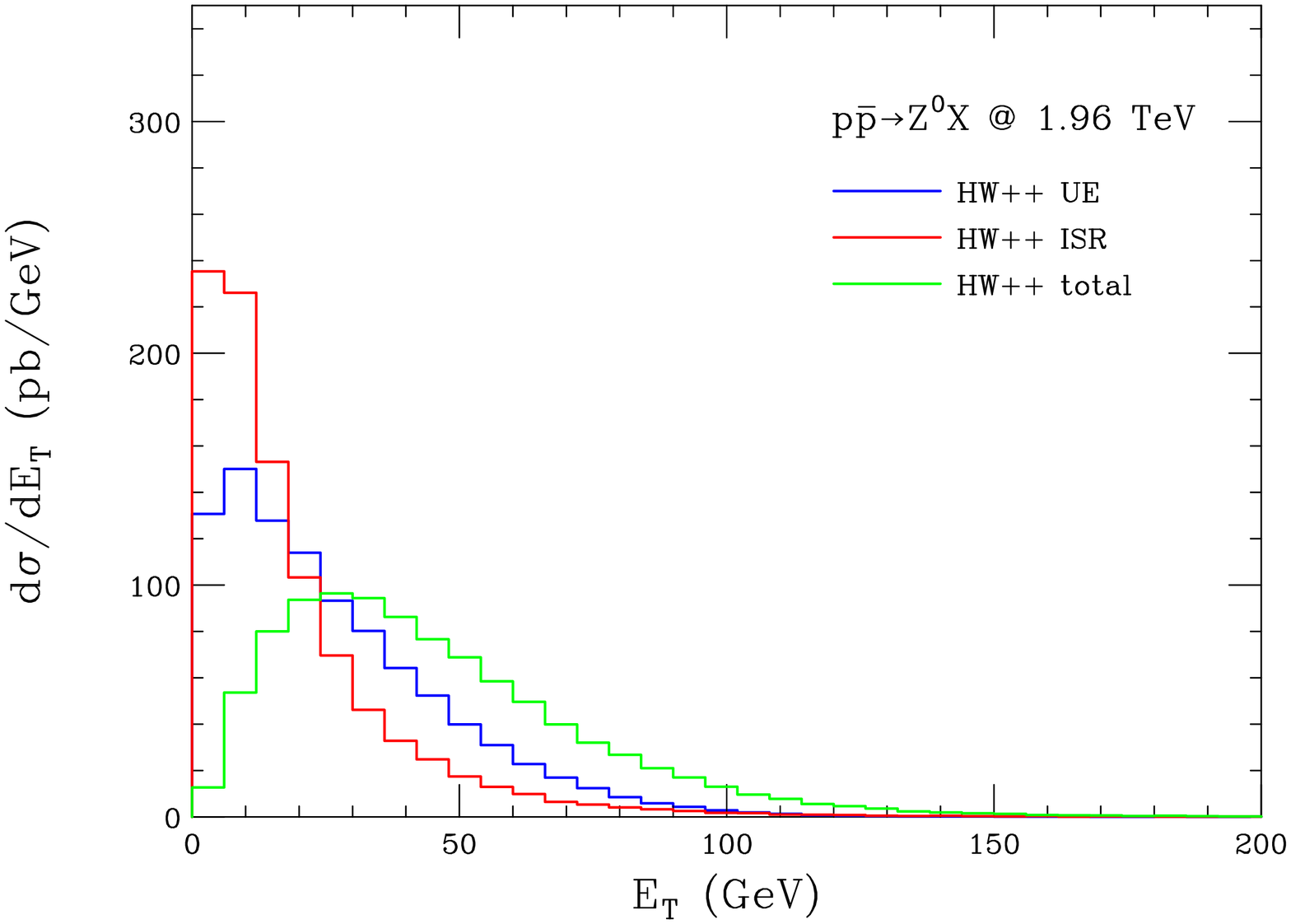,angle=90,width=75mm}
\epsfig{file=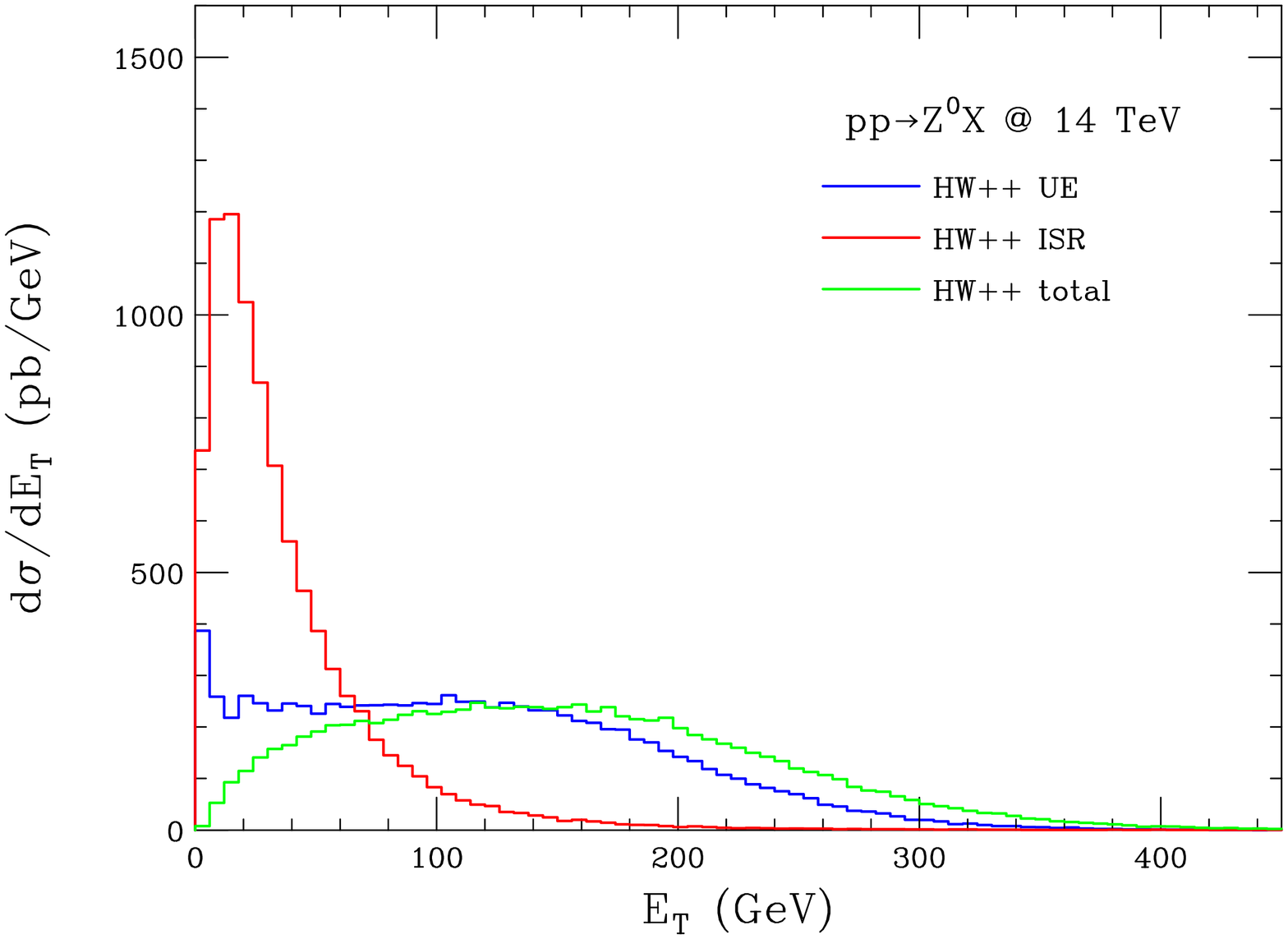,angle=90,width=75mm}
\end{center}
\caption{Predicted $E_T$ distribution in Z$^0$ boson production  at the
 Tevatron and LHC. Monte Carlo results including underlying event.
\label{fig:ZTevall} }
\end{figure}

\begin{figure}
\begin{center}
\epsfig{file=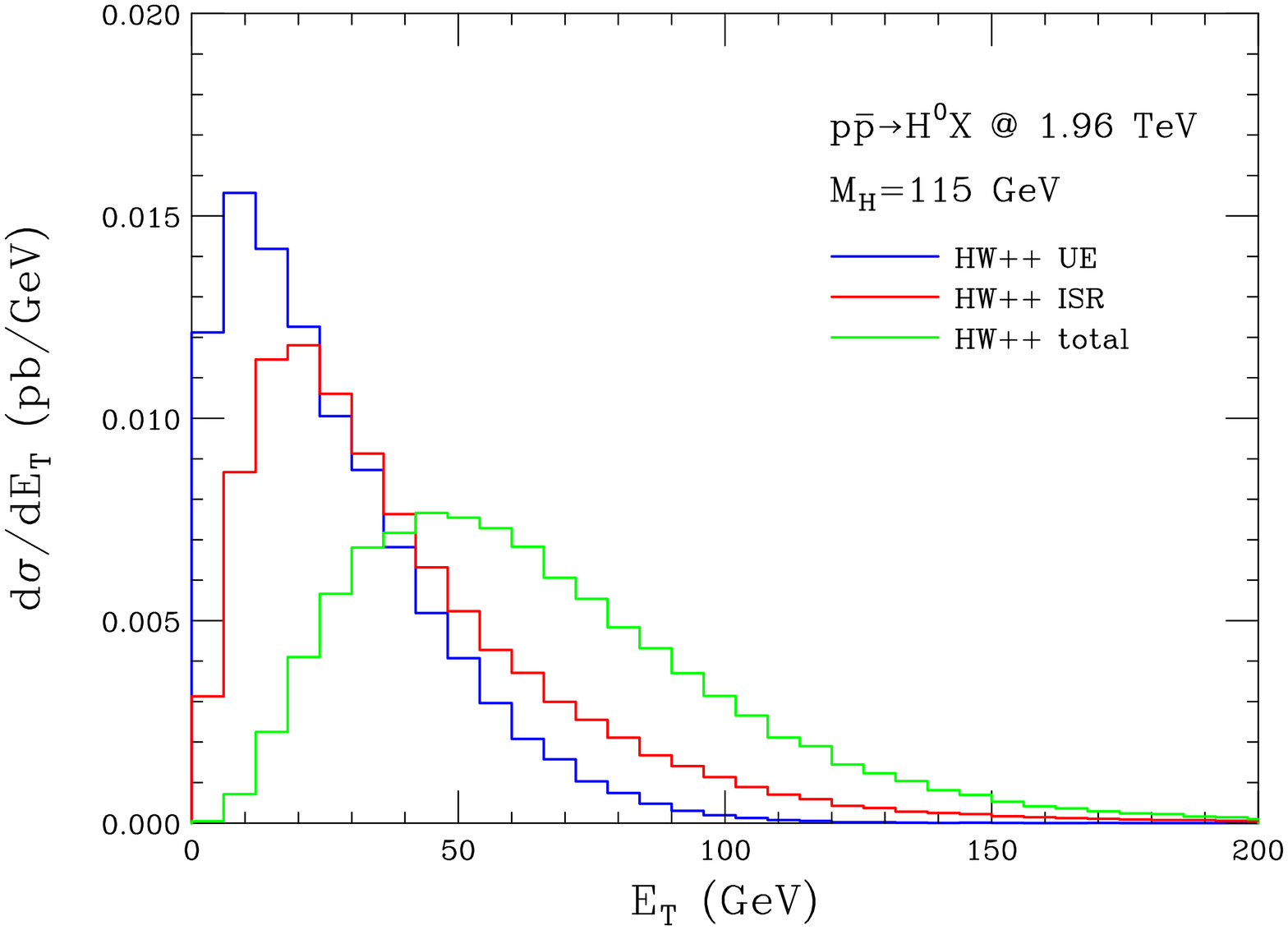,angle=90,width=75mm}
\epsfig{file=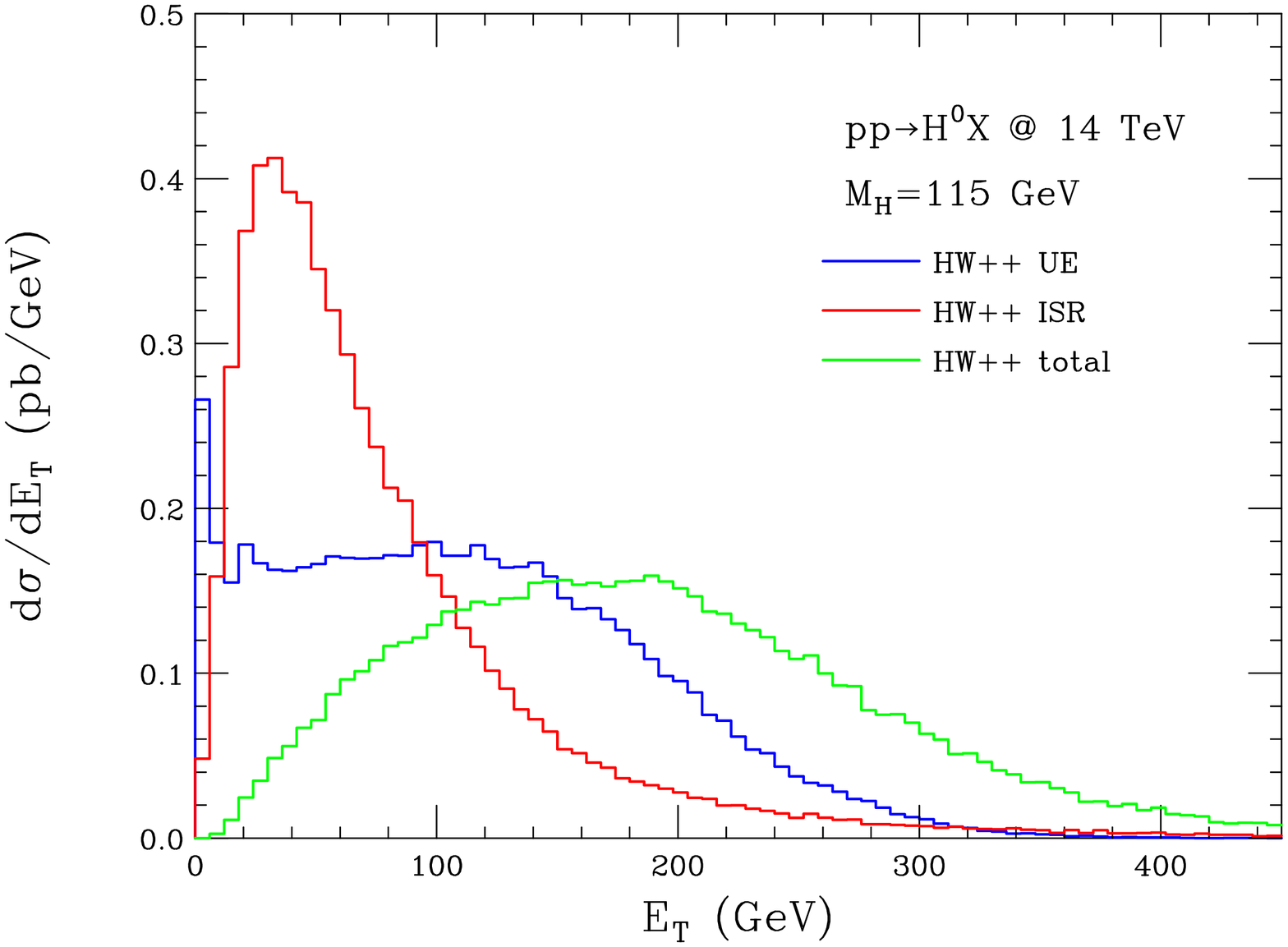,angle=90,width=75mm}
\end{center}
\caption{Predicted $E_T$ distribution in Higgs boson production  at the
 Tevatron and LHC. Monte Carlo results including underlying event.
\label{fig:HTevall} }
\end{figure}

\begin{figure}
\begin{center}
\epsfig{file=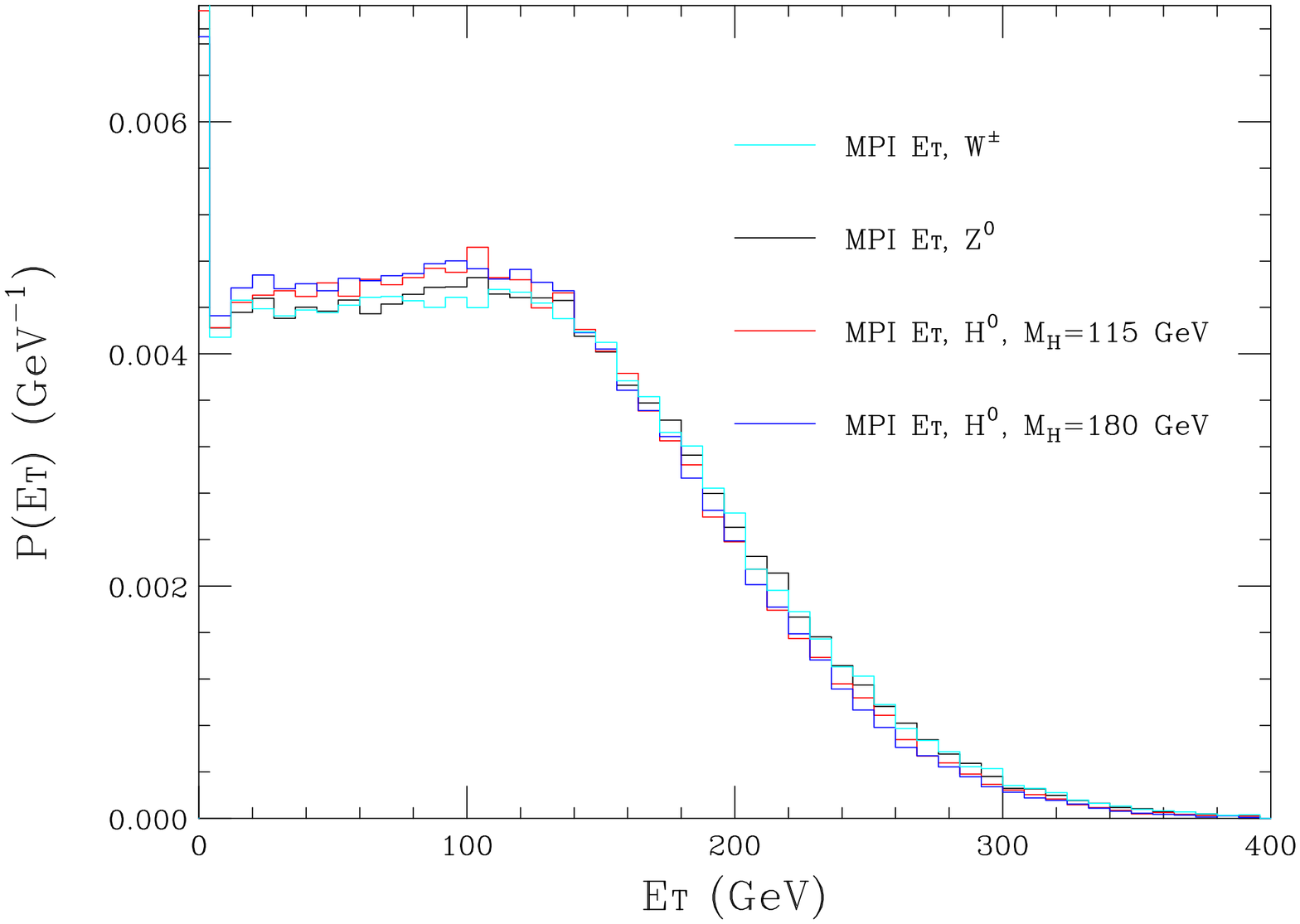,angle=90,width=125mm}
\end{center}
\caption{Comparison of $E_T$ distributions of the underlying event
in different subprocesses at the LHC.
\label{fig:MPIcomp} }
\end{figure}

Figures~\ref{fig:ZTevall} and \ref{fig:HTevall}  show the parton-level  {\tt Herwig++}
predictions for the $E_T$ distribution in Z$^0$ and Higgs boson
production, respectively, with the contributions from initial-state
radiation (in red, already shown in Figs.~\ref{fig:ZfHWall}
and \ref{fig:HfHWall} ), the underlying event (blue) and the
combination of the two (green).  The underlying event is modelled
using multiple parton interactions; see ref.~\cite{Bahr:2008pv} for details.
Clearly it has a very significant effect on the $E_T$ distribution.
However, this effect is substantially independent of the hard
subprocess, as may be seen from the comparison of different
subprocesses in Fig.~\ref{fig:MPIcomp}.

\begin{figure}
\begin{center}
\epsfig{file=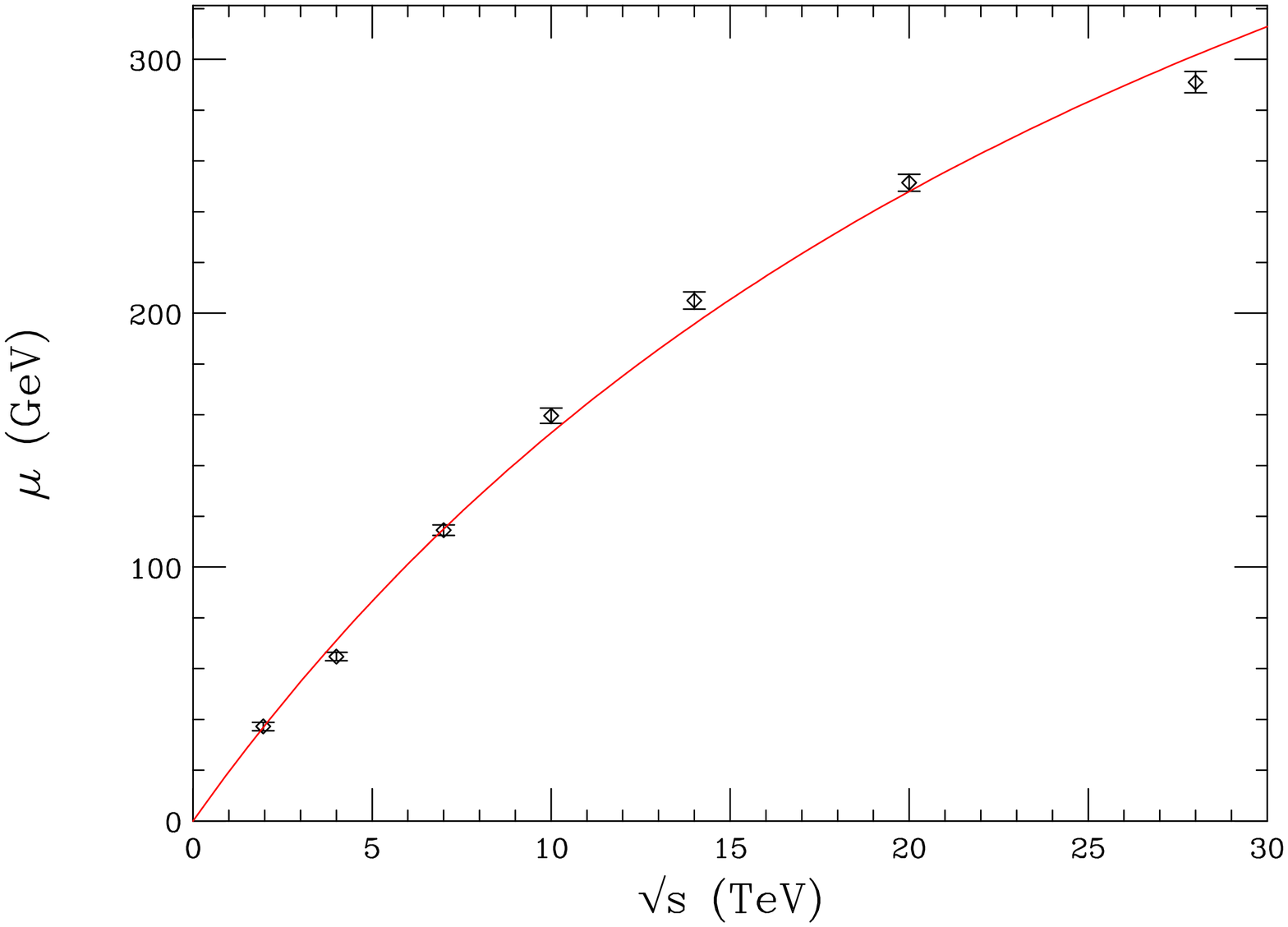,angle=90,width=75mm}
\epsfig{file=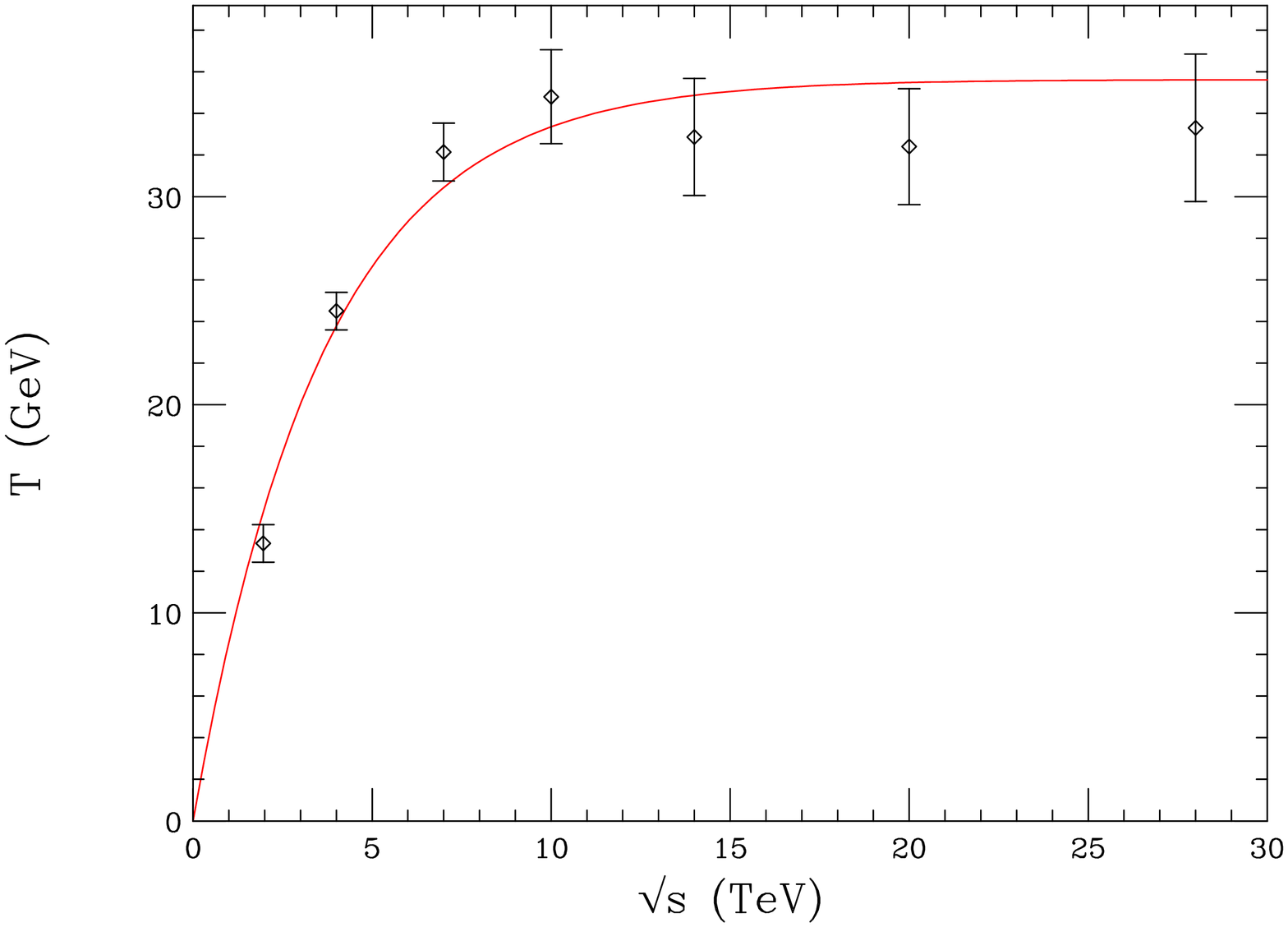,angle=90,width=75mm}
\end{center}
\caption{Fitted values of the parameters of the underlying event in
  Higgs production in $pp$ collisions at various energies.
\label{fig:MPIfits} }
\end{figure}

We find that the probability distribution of the $E_T$ contribution of
the underlying event in the {\tt  Herwig++} Monte Carlo can be
represented quite well by a Fermi distribution:
\beq\label{eq:Fermi}
P(E_T) = \frac 1{\cal N}\frac 1{\exp\left(\frac{E_T-\mu}T\right)+1}
\eeq
where the normalization is
\beq
{\cal N} = T\ln\left[\exp\left(\frac{\mu}T\right)+1\right]\;.
\eeq
The dependence of the ``chemical potential'' $\mu$ and ``temperature''
$T$ on the hadronic collision energy is shown in
Fig.~\ref{fig:MPIfits}.  The red curves show fits to the energy
dependence of the form
\beq
\mu =\frac{A\rs}{1+B\rs}\,,\quad
T = q\left(1-{\rm e}^{-r\rs}\right)
\eeq
where the coefficients in the fits are $A = 20(1)$, $B = 0.030(4)$,
$q = 36(2)$, $r = 0.28(3)$.

\section{Conclusions}\label{sec:conc}
 We have extended the resummation of the hadronic transverse energy $E_T$ in vector
boson production to next-to-leading order (NLO) in the resummed exponent,
parton distributions and coefficient functions, and also presented for the
first time the corresponding predictions for Higgs boson production.
We have matched the resummed results to the corresponding ${\cal O}(\as)$
predictions, by adding the contributions in that order which are not included
in the resummation.  In addition we have compared with parton shower Monte Carlo
results and illustrated the effects of hadronization and the
underlying event.

In the case of vector boson production, the resummation procedure
appears stable and the parton-level results should be quite reliable.
The leading-order mechanism of quark-antiquark annihilation typically
generates a moderate amount of transverse energy in initial-state QCD
radiation.  Consequently  the effects of subleading resummed terms and
fixed-order matching are small and the peak of the $E_T$ distribution
lies well below the boson mass scale, where resummation makes good
sense.  The comparisons with Monte Carlo programs reveal some
discrepancies but these are at the level of disagreements between
different programs; in this case the resummed predictions should be
more reliable (at parton level) than existing Monte Carlos.  The
programs suggest that the non-perturbative effects of hadronization
and the underlying event are substantial.  These effects can however
be modelled in a process-independent way.  We have suggested a
simple parametrization of the contribution of the underlying event.

The situation in Higgs boson production is not so good.  The dominant
mechanism of gluon fusion generates copious ISR and the effects of
subleading terms and matching are large.  The resummed $E_T$
distribution peaks at a value that is not parametrically smaller than
the Higgs mass and the behaviour at low and high $E_T$ is unphysical
before matching. The discrepancies between the matched resummed and
Monte Carlo predictions are substantially greater than those between
different programs, even allowing for uncertainties in the overall
cross section. All this suggests that there are significant higher-order
corrections that are not taken into account, either further subleading
logarithms or unenhanced terms beyond NLO.  It would be
interesting (but very challenging) to attempt to extract such terms
from the available NNLO calculations of Higgs production.

\section*{Acknowledgements}
We are grateful for helpful correspondence and discussions with
Stefano Catani and  James Stirling.  JS and BW thank the CERN Theory
Group for hospitality during part of this work.  This work was
supported in part by the UK Science and Technology Facilities Council
and the European Union Marie Curie Research Training Network MCnet
(contract MRTN-CT-2006-035606).

\appendix
\section{Relation to transverse momentum resummation}
Here we demonstrate the equivalence of transverse energy and transverse momentum resummation at order $\as$.  Expanding Eq.~(\ref{formfact}) to this order, using (\ref{eq:nllint}) and substituting into (\ref{resgen}) and (\ref{eq:Wab}), we find terms involving the integrals
\beq\label{eq:Ip}
{\cal I}_p(Q,E_T) = \frac 1{2\pi}\int_{-\infty}^{+\infty} d\tau \; {\rm e}^{-i\tau E_T} 
\ln^p\left(\frac{Q\tau}{i\tau_0}\right)
\eeq
with $p=1,2$.  At this order, evaluating the PDFs at the scale $i\tau_0/\tau$ leads to single-logarithmic terms of the same form when we use (\ref{eq:fabi}) to write
\beq\label{eq:fQ}
f_{a/h}(x,i\tau_0/\tau)) = f_{a/h}(x,Q) -\frac{\as}{\pi} \ln\left(\frac{Q\tau}{i\tau_0}\right)\sum_b \int_x^1 \frac{dz}z P_{ab}(z) f_{b/h}(x/z,Q)\;.
\eeq

The integral (\ref{eq:Ip}) may be evaluated from
\beq\label{eq:IpIu}
{\cal I}_p(Q,E_T) = \frac{d^p}{du^p}{\cal I}(Q,E_T;u)|_{u=0}
\eeq
where
\beq\label{eq:Iu}
{\cal I}(Q,E_T;u) = \frac 1{2\pi}\int_{-\infty}^{+\infty} d\tau \; {\rm e}^{-i\tau E_T}\;.
\left(\frac{Q\tau}{i\tau_0}\right)^u
\eeq
Writing $\tau=iz/E_T$, we have
\beq
{\cal I}(Q,E_T;u) = -\frac i{2\pi E_T}\left(\frac Q{E_T\tau_0}\right)^u
\int_{-i\infty}^{+i\infty} dz\; z^u\,{\rm e}^z\;.
\eeq
We can safely deform the integration contour around the branch cut along the negative real axis to obtain
\beq\label{eq:Iufin}
{\cal I}(Q,E_T;u) = -\frac 1{\pi E_T}\left(\frac Q{E_T\tau_0}\right)^u
\sin(\pi u)\,\Gamma(1+u)\;,
\eeq
which, recalling that $\ln\tau_0 =-\gamma_E = \Gamma'(1)$, gives
\beq\label{eq:I12}
{\cal I}_1(Q,E_T) = -\frac 1{E_T}\;,\quad
{\cal I}_2(Q,E_T) = -\frac 2{E_T}\ln\left(\frac Q{E_T}\right)\;.
\eeq

The resummed component of the transverse momentum ($q_T$) distribution takes the form
\beeq
\label{eq:qtres}
\left[ \frac{d\sigma_{F}}{dQ^2\;dq_T} \right]_{\res} &=& q_T\sum_{a,b}
\int_0^1 dx_1 \int_0^1 dx_2 \int_0^{\infty} db\,b \,J_0(bq_T)
\;f_{a/h_1}(x_1,b_0/b) \; f_{b/h_2}(x_2,b_0/b) \nn \\
&\cdot& \overline W_{ab}^{F}(x_1 x_2 s; Q,b)
\eeeq
where $b_0=2\exp(-\gamma_E)$,
\beeq 
\label{eq:qtWab}
\overline W_{ab}^{F}(s; Q,b) &=& \sum_c \int_0^1 dz_1 \int_0^1 dz_2 
\; C_{ca}(\as(b_0/b), z_1) \; C_{{\bar c}b}(\as(b_0/b), z_2)
\; \delta(Q^2 - z_1 z_2 s) \nn\\
&\cdot& \sigma_{c{\bar c}}^F(Q,\as(Q)) \;\overline S_c(Q,b)
\eeeq
and
\beq
\label{eq:qtff}
\overline S_c(Q,b) = \exp \left\{-2\int_{b_0/b}^Q \frac{dq}q 
\left[ 2A_c(\as(q)) \;\ln \frac{Q}{q} + B_c(\as(q)) \right]\right\} \;.
\eeq
Expanding to order $\as$, we find the same terms as in the $E_T$ resummation except that (\ref{eq:Ip}) is replaced by
\beq\label{eq:qTIp}
\overline {\cal I}_p(Q,q_T) =  q_T\int_0^{\infty} db\,b \,J_0(bq_T)\ln^p(Qb/b_0)\;.
\eeq
It therefore suffices to show that
\beq\label{eq:qTIpIp}
\overline {\cal I}_p(Q,q_T) =  {\cal I}_p(Q,E_T=q_T)\;\;\;\mbox{for  $p=1,2$.}
\eeq
Now corresponding to (\ref{eq:Iu}) we have
\beq\label{eq:qTIu}
\overline {\cal I}(Q,q_T;u) =  q_T\int_0^{\infty} db\,b \,J_0(bq_T)\left(\frac{Qb}{b_0}\right)^u\;.
\eeq
Using the result
\beq
\int_0^{\infty} dt\,t^{\mu-1} \,J_0(t) = \frac{2^\mu}{2\pi}\sin\left(\frac{\pi\mu}2\right)
\Gamma^2\left(\frac{\mu}2\right)
\eeq
gives
\beq
\overline {\cal I}(Q,q_T;u) =-\frac{2}{\pi q_T}\left(\frac{2Q}{q_T b_0}\right)^u\sin\left(\frac{\pi u}2\right)
\Gamma^2\left(1+\frac u2\right)
\eeq
and hence
\beq
\overline {\cal I}_1(Q,q_T) = -\frac 1{q_T}\;,\quad
\overline {\cal I}_2(Q,q_T) = -\frac 2{q_T}\ln\left(\frac Q{q_T}\right)\;,
\eeq
in agreement with  (\ref{eq:I12}) and (\ref{eq:qTIpIp}).  Notice, however, that the higher ($p>2$)
derivatives of ${\cal I}$ and $\overline {\cal I}$ differ, corresponding to the difference between $E_T$ and
$q_T$ resummation beyond ${\cal O}(\as)$.

\newpage
\section{Results for the LHC at 7 TeV}
\begin{figure}[h]
\begin{center}
\epsfig{file=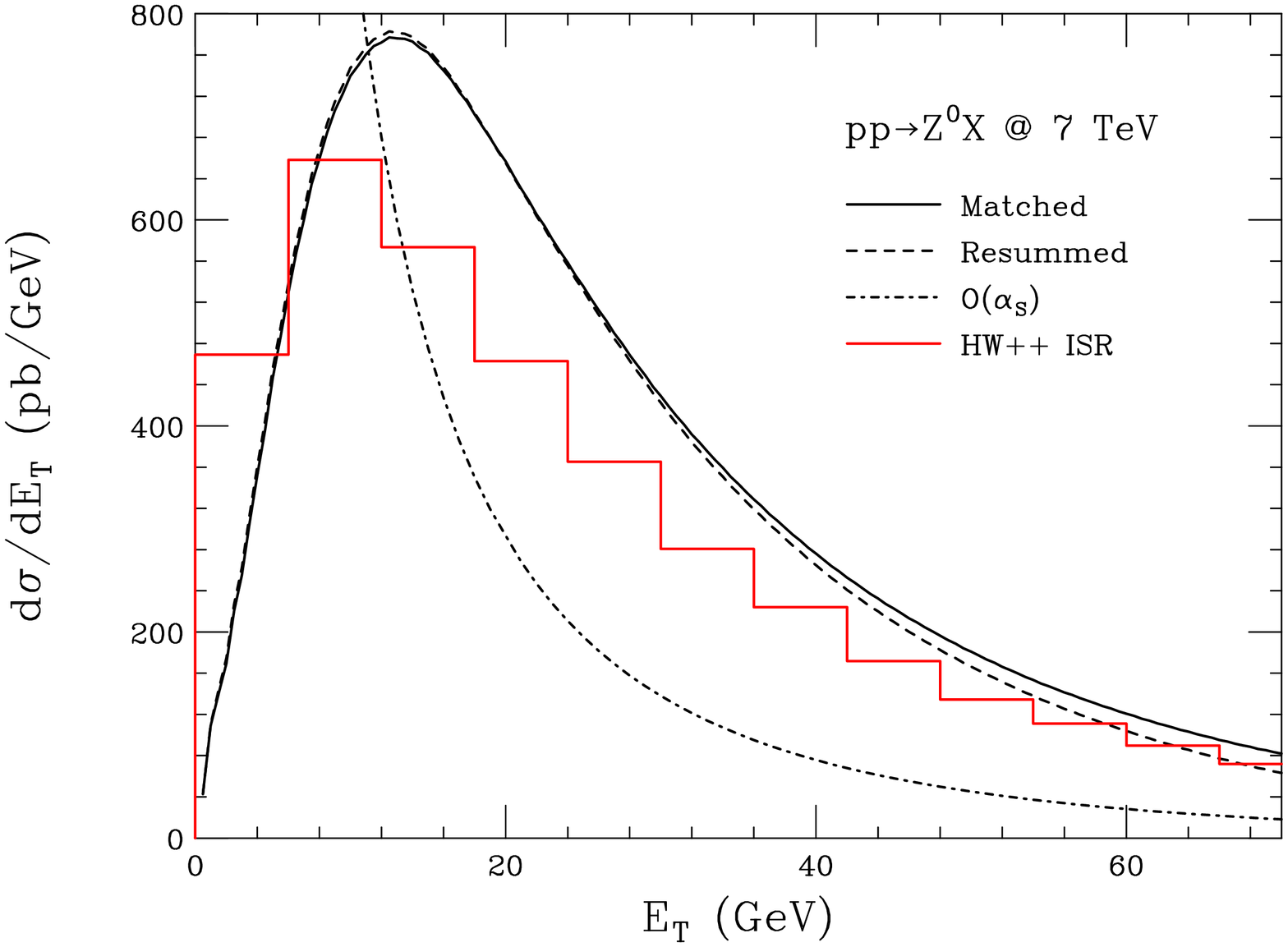,angle=90,width=75mm}
\epsfig{file=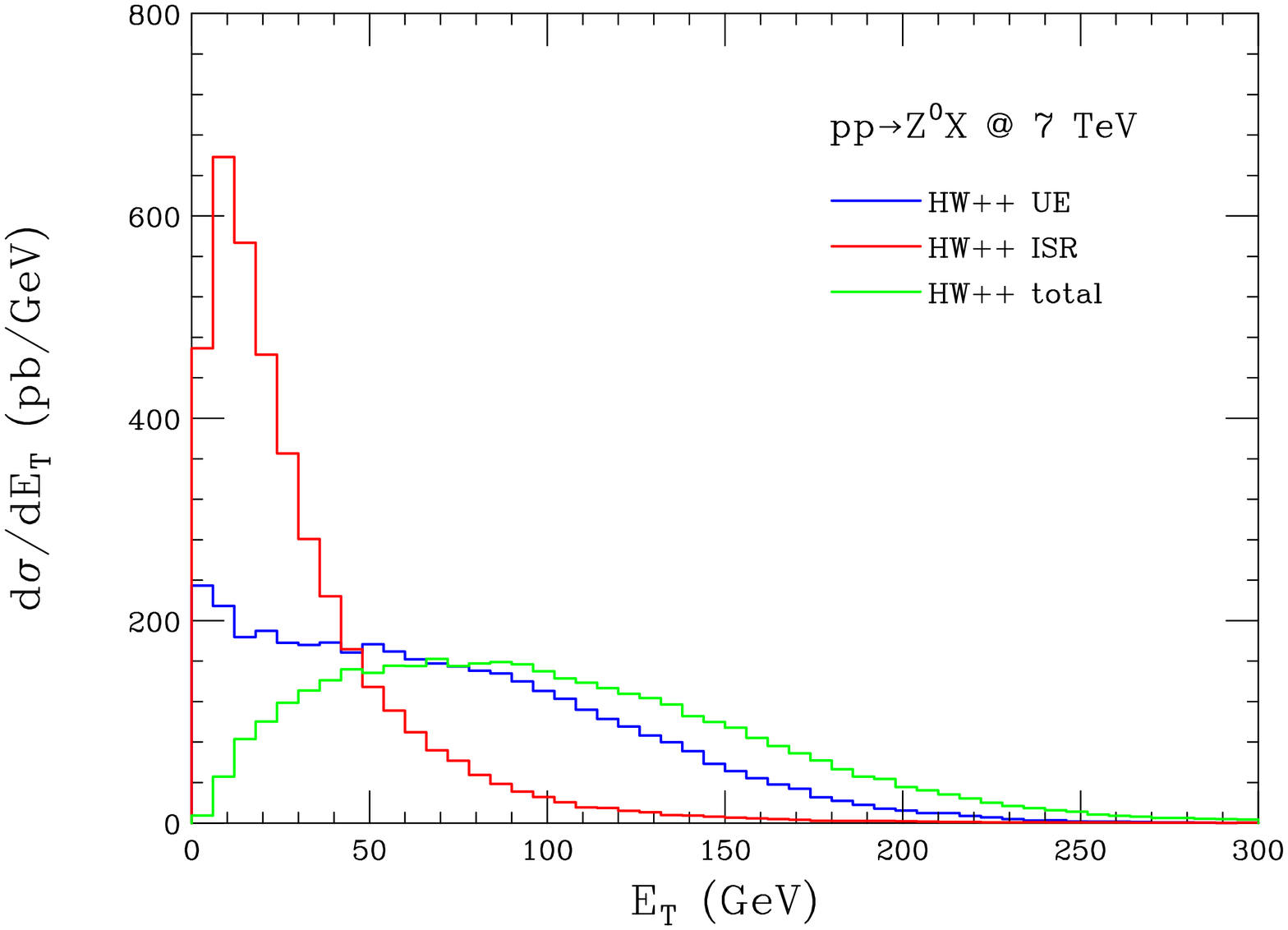,angle=90,width=75mm}
\epsfig{file=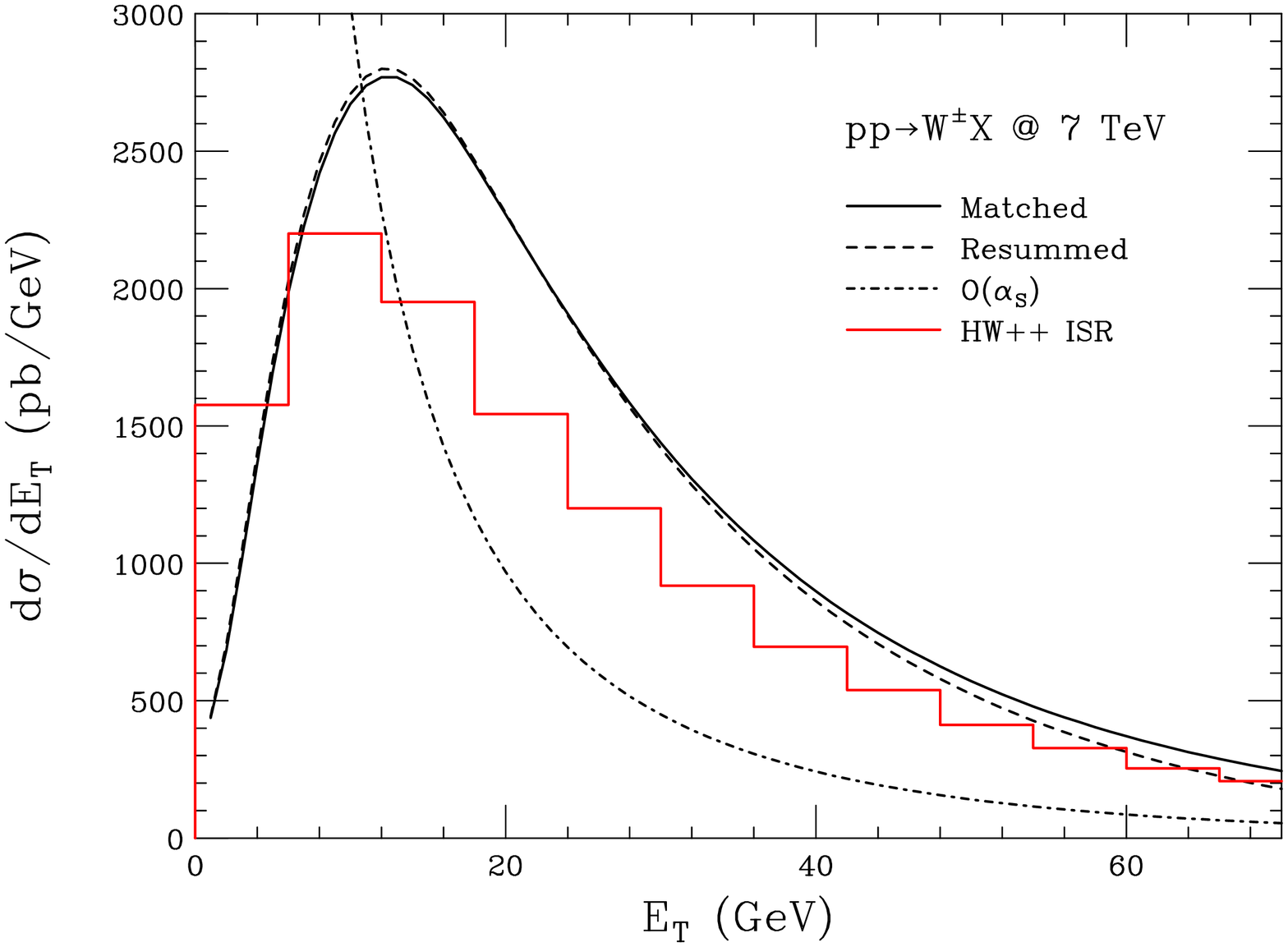,angle=90,width=75mm}
\epsfig{file=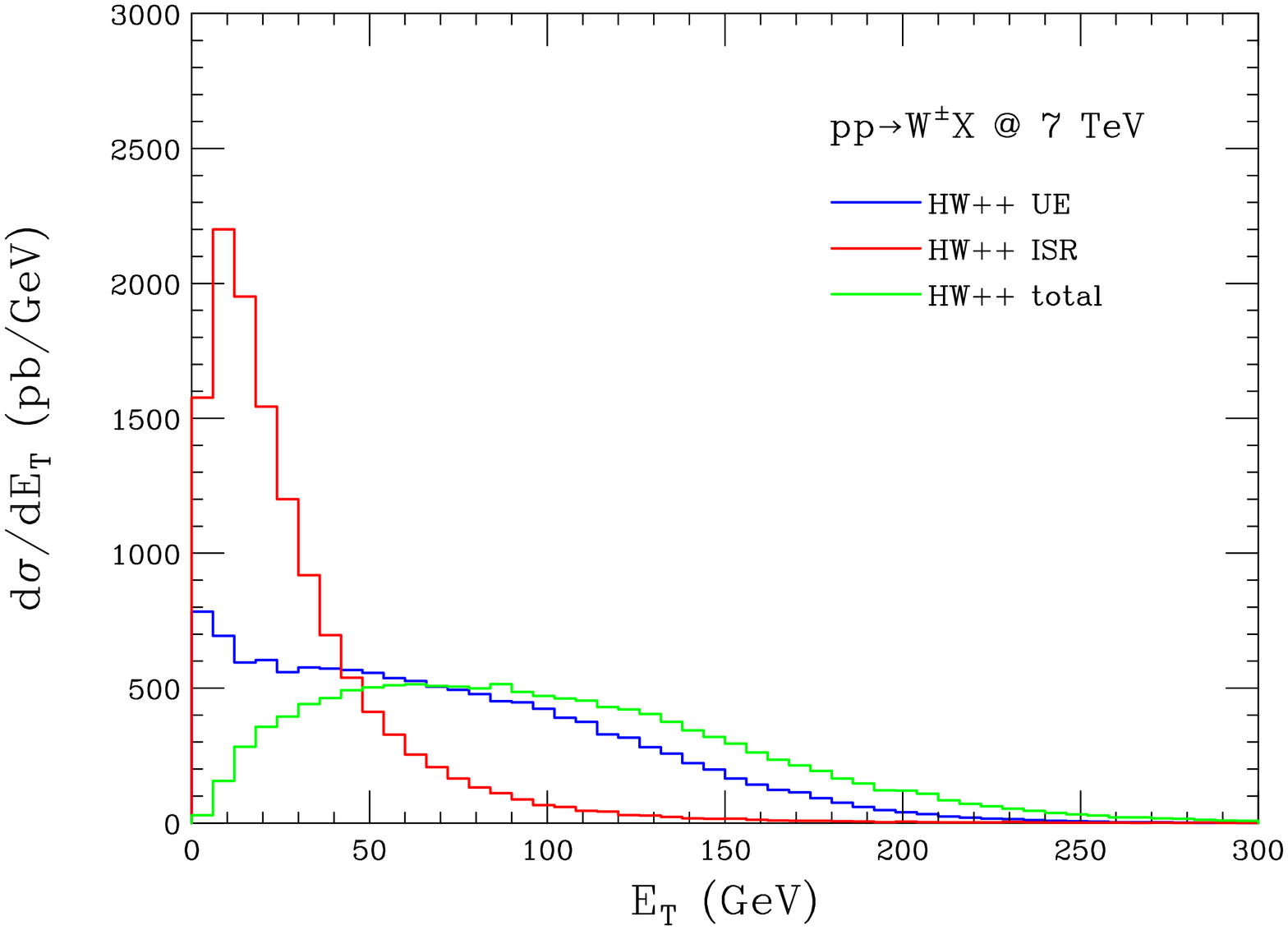,angle=90,width=75mm}
\epsfig{file=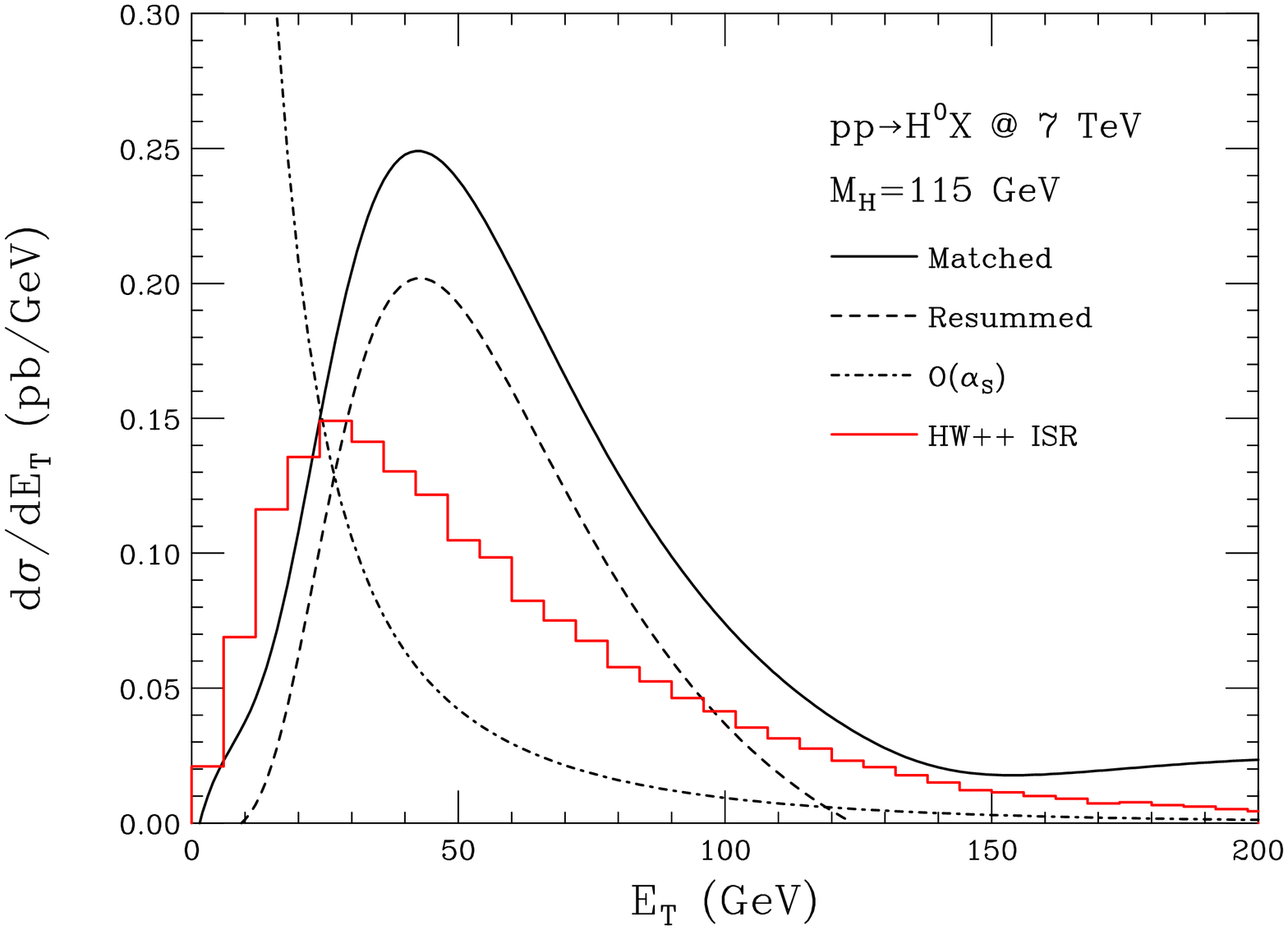,angle=90,width=75mm}
\epsfig{file=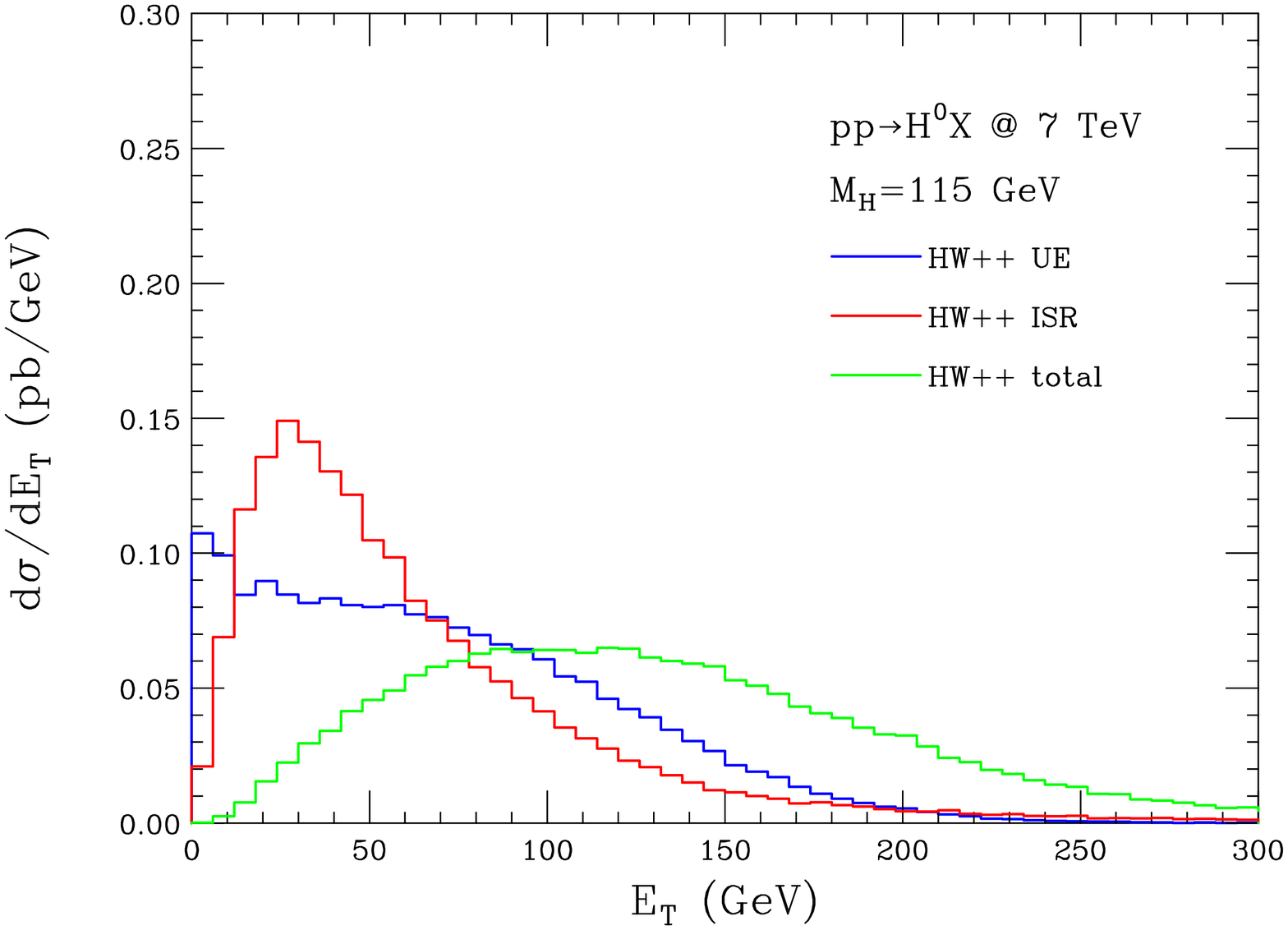,angle=90,width=75mm}
\end{center}
\caption{Predicted $E_T$ distributions in Z$^0$,   W$^+$+W$^-$ and
  Higgs boson production in $pp$ collisions at $\rs =7$ TeV .
\label{fig:LHC7all} }
\end{figure}
We show here results for the LHC operating at a centre-of-mass energy
of 7 TeV, corresponding to those shown earlier for 14 TeV.  Apart from
the normalization, the predictions for the two energies are very
similar, with only a slight downward shift in the position of the peak in the
$E_T$ distribution at the lower energy.



\begin{thebibliography}{999}

\bibitem{Bozzi:2007pn}
  G.~Bozzi, S.~Catani, D.~de Florian and M.~Grazzini,
  Nucl.\ Phys.\  B {\bf 791} (2008) 1
  [arXiv:0705.3887 [hep-ph]].

\bibitem{Bozzi:2008bb}
  G.~Bozzi, S.~Catani, G.~Ferrera, D.~de Florian and M.~Grazzini,
  Nucl.\ Phys.\  B {\bf 815}, 174 (2009)
  [arXiv:0812.2862 [hep-ph]].

\bibitem{Mantry:2009qz}
  S.~Mantry and F.~Petriello,
  arXiv:0911.4135.

\bibitem{Dokshitzer:1980hw}
Y.~L.~Dokshitzer, D.~Diakonov and S.~I.~Troian,
Phys.\ Rep.\  58 (1980) 269.

\bibitem{Parisi:1979se}
G.~Parisi and R.~Petronzio,
Nucl.\ Phys.\  B154 (1979) 427.

\bibitem{Curci:1979bg}
G.~Curci, M.~Greco and Y.~Srivastava,
Nucl.\ Phys.\  B159 (1979) 451.

\bibitem{Bassetto:1980nt}
A.~Bassetto, M.~Ciafaloni and G.~Marchesini,
Nucl.\ Phys.\  B163 (1980) 477.

\bibitem{Kodaira:1982nh}
J.~Kodaira and L.~Trentadue,
Phys.\ Lett.\  B112 (1982) 66,
Phys.\ Lett.\  B123 (1983) 335.

\bibitem{Collins:1984kg}
  J.~C.~Collins, D.~E.~Soper and G.~Sterman,
  Nucl.\ Phys.\  B {\bf 250} (1985) 199.

\bibitem{Halzen:1982cb}
  F.~Halzen, A.~D.~Martin, D.~M.~Scott and M.~P.~Tuite,
  Z.\ Phys.\  C {\bf 14} (1982) 351.

\bibitem{Davies:1983di}
  C.~T.~H.~Davies and B.~R.~Webber,
  Z.\ Phys.\  C {\bf 24} (1984) 133.

\bibitem{Altarelli:1986zk}
  G.~Altarelli, G.~Martinelli and F.~Rapuano,
  Z.\ Phys.\  C {\bf 32} (1986) 369.

\bibitem{Corcella:2000bw}
  G.~Corcella, I.~G.~Knowles, G.~Marchesini, S.~Moretti, K.~Odagiri, P.~Richardson, M.~H.~Seymour and B.~R.~Webber,
  JHEP {\bf 0101} (2001) 010
  [arXiv:hep-ph/0011363]; arXiv:hep-ph/0210213.
{\tt http://projects.hepforge.org/fherwig/}

\bibitem{Bahr:2008pv}
  M.~Bahr {\it et al.},
  Eur.\ Phys.\ J.\  C {\bf 58} (2008) 639
  [arXiv:0803.0883 [hep-ph]];
  arXiv:0812.0529 [hep-ph].
{\tt http://projects.hepforge.org/herwig/}

\bibitem{Catani:2000vq}
  S.~Catani, D.~de Florian and M.~Grazzini,
  Nucl.\ Phys.\  B {\bf 596}, 299 (2001)
  [arXiv:hep-ph/0008184].

\bibitem{Davies:1984hs}
C.~T.~Davies and W.~J.~Stirling,
Nucl.\ Phys.\  B244 (1984) 337.

\bibitem{Balazs:1995nz}
 C.~Balazs, J.~W.~Qiu and C.~P.~Yuan,
 Phys.\ Lett.\  B {\bf 355} (1995) 548
 [arXiv:hep-ph/9505203].

\bibitem{deFlorian:2000pr}
  D.~de Florian and M.~Grazzini,
  Phys.\ Rev.\ Lett.\  {\bf 85} (2000) 4678
  [arXiv:hep-ph/0008152].

\bibitem{deFlorian:2001zd}
  D.~de Florian and M.~Grazzini,
  Nucl.\ Phys.\  B {\bf 616} (2001) 247
  [arXiv:hep-ph/0108273].

\bibitem{Catani:1988vd}
S.~Catani, E.~D'Emilio and L.~Trentadue,
Phys.\ Lett.\  B211 (1988) 335.

\bibitem{Kauffman:1992cx}
R.~P.~Kauffman,
Phys.\ Rev.\  D45 (1992) 1512.

\bibitem{Martin:2009iq}
  A.~D.~Martin, W.~J.~Stirling, R.~S.~Thorne and G.~Watt,
  Eur.\ Phys.\ J.\  C {\bf 63} (2009) 189
  [arXiv:0901.0002 [hep-ph]].

\bibitem{Butterworth:1996zw}
  J.~M.~Butterworth, J.~R.~Forshaw and M.~H.~Seymour,
  Z.\ Phys.\  C {\bf 72} (1996) 637
  [arXiv:hep-ph/9601371].
{\tt http://projects.hepforge.org/jimmy/}

\bibitem{Sherstnev:2007nd}
  A.~Sherstnev and R.~S.~Thorne,
  Eur.\ Phys.\ J.\  C {\bf 55} (2008) 553
  [arXiv:0711.2473 [hep-ph]];
  arXiv:0807.2132 [hep-ph].

\end{thebibliography}
\end{document}